\documentclass[modern]{aastex631}

\usepackage{lineno}

\newcommand{\kms}{${\rm km~s}^{-1}$}

\newcommand{\dm}{$\Delta{\rm m}_{15}(B)$}
\newcommand{\sbv}{$s_{BV}$}

\submitjournal{ApJ}

\begin{document}

\title{The Absolute Magnitudes of 1991T-like Supernovae\footnote{This paper includes data gathered with the 6.5 meter 
Magellan telescopes at Las Campanas Observatory, Chile.}}

\author[0000-0003-2734-0796]{M.~M.~Phillips}
\affiliation{Carnegie Observatories, Las Campanas Observatory, Casilla 601, La Serena, Chile}

\author[0000-0002-5221-7557]{C.~Ashall}
\affiliation{Department of Physics, Virginia Polytechnic Institute and State University, 850 West Campus Drive, Blacksburg, VA 24061, USA}

\author[0000-0003-4625-6629]{Christopher R.~Burns}
\affiliation{Observatories of the Carnegie Institution for Science, 813 Santa Barbara St., Pasadena, CA 91101, USA}

\author[0000-0001-6293-9062]{Carlos~Contreras}
\affiliation{Carnegie Observatories, Las Campanas Observatory, Casilla 601, La Serena, Chile}

\author[0000-0002-1296-6887]{L.~Galbany}
\affiliation{Institute of Space Sciences (ICE, CSIC), Campus UAB, Carrer de Can Magrans, s/n, E-08193 Barcelona, Spain.}
\affiliation{Institut d'Estudis Espacials de Catalunya (IEEC), E-08034 Barcelona, Spain.}

\author[0000-0002-4338-6586]{P.~Hoeflich}
\affiliation{Department of Physics, Florida State University, 77 Chieftan Way, Tallahassee, FL  32306, USA}

\author[0000-0003-1039-2928]{E.~Y.~Hsiao}
\affiliation{Department of Physics, Florida State University, 77 Chieftan Way, Tallahassee, FL  32306, USA}

\author[0000-0003-2535-3091]{Nidia~Morrell}
\affiliation{Carnegie Observatories, Las Campanas Observatory, Casilla 601, La Serena, Chile}

\author[0000-0002-3389-0586]{Peter Nugent}
\affiliation{Department of Astronomy, University of California, Berkeley, USA}
\affiliation{Lawrence Berkeley National Laboratory, USA}

\author[0000-0002-9413-4186]{Syed~A.~Uddin}
\affiliation{George P. and Cynthia Woods Mitchell Institute for Fundamental Physics and Astronomy, Texas A\&M University,
Department of Physics and Astronomy,  College Station, TX 77843, USA}

\author[0000-0001-5393-1608]{E. Baron}
\affiliation{University of Oklahoma 440 W. Brooks, Rm 100, Norman, Oklahoma, 73019, USA}

\author[0000-0003-3431-9135]{Wendy~L.~Freedman}
\affiliation{Department of Astronomy and Astrophysics, University of Chicago, 5640 S. Ellis Ave, Chicago, IL 60637, USA}

\author[0000-0002-1751-7474]{Chelsea E. Harris}
\affiliation{Center for Data Intensive and Time Domain Astronomy, Department of Physics and Astronomy, Michigan State University, East Lansing, MI 48824, USA}

\author[0000-0002-6650-694X]{Kevin~Krisciunas}
\affiliation{George P. and Cynthia Woods Mitchell Institute for Fundamental Physics and Astronomy, Texas A\&M University,
Department of Physics and Astronomy,  College Station, TX 77843, USA}

\author[0000-0001-8367-7591]{S. Kumar}
\affiliation{Department of Physics, Florida State University, 77 Chieftan Way, Tallahassee, FL  32306, USA}

\author[0000-0002-3900-1452]{J. Lu}
\affiliation{Department of Physics, Florida State University, 77 Chieftan Way, Tallahassee, FL  32306, USA}

\author[0000-0003-0554-7083]{S.~E.~Persson}
\affiliation{Observatories of the Carnegie Institution for Science, 813 Santa Barbara St., Pasadena, CA 91101, USA}

\author[0000-0001-6806-0673]{Anthony L. Piro}
\affiliation{Observatories of the Carnegie Institution for Science, 813 Santa Barbara St., Pasadena, CA 91101, USA}

\author[0000-0002-1633-6495]{Abigail Polin}
\affiliation{Observatories of the Carnegie Institution for Science, 813 Santa Barbara St., Pasadena, CA 91101, USA}
\affiliation{TAPIR, Walter Burke Institute for Theoretical Physics, Caltech, 1200 East California Boulevard, Pasadena, CA 91125, USA}

\author[0000-0002-9301-5302]{Shahbandeh, M.}
\affiliation{Department of Physics, Florida State University, 77 Chieftan Way, Tallahassee, FL  32306, USA}

\author[0000-0002-5571-1833]{Maximilian Stritzinger}
\affiliation{Department of Physics and Astronomy, Aarhus University, Ny Munkegade 120, DK-8000 Aarhus C, Denmark}

\author[0000-0002-8102-181X]{Nicholas~B.~Suntzeff}
\affiliation{George P. and Cynthia Woods Mitchell Institute for Fundamental Physics and Astronomy, Texas A\&M University,
Department of Physics and Astronomy,  College Station, TX 77843, USA}



\begin{abstract}

1991T-like supernovae are the luminous, slow-declining extreme of the Branch shallow-silicon (SS) subclass of Type~Ia supernovae.  They are
distinguished by extremely weak \ion{Ca}{2}~H \& K and \ion{Si}{2}~$\lambda6355$ and strong \ion{Fe}{3} absorption features in their optical spectra at pre-maximum phases,
and have long been suspected to be over-luminous compared to normal Type~Ia supernovae.
In this paper, the pseudo equivalent width of the \ion{Si}{2}~$\lambda$6355 absorption obtained at light curve 
phases from $\leq+10$~days is combined with the morphology of the $i$-band light curve to identify a sample of 1991T-like supernovae
in the Carnegie Supernova Project~II.  Hubble diagram 
residuals show that, at optical as well as near-infrared wavelengths, these events are 
over-luminous by $\sim$0.1--0.5~mag with respect to the
less extreme Branch SS 
(1999aa-like) and Branch core-normal supernovae with similar
$B$-band light curve decline rates.

\end{abstract}



\section{Introduction}
\label{sec:intro}

Type~Ia supernovae (SNe~Ia) provide a powerful tool for modern cosmology.  Although they are not perfect standard candles, the 
luminosities and colors of SNe~Ia vary smoothly with light curve width \citep{phillips93,hamuy96,riess96,phillips99,burns14,burns18} 
making it possible to determine reliable distances to a precision of 5--10\%, especially when near-infrared (NIR) photometry is
available \citep[e.g.,][]{krisciunas04,phillips12,burns18,avelino19}.  Observations of SNe~Ia have provided the most precise local measurements of the 
Hubble constant to date \citep[e.g.,][]{riess16,dhawan18,burns18,freedman19,riess01,uddin22} and led to the discovery of the accelerated expansion of the 
Universe \citep{riess98,perlmutter99}.

At the luminous extreme of the absolute magnitude versus light-curve width relation for SNe~Ia are the slow-declining 1991T-like
(henceforth ``91T-like'') events, named 
after the well-observed SN~1991T \citep{filippenko92,phillips92,ruiz-lapuente92,jeffery92,mazzali95,gomez96,lira98,blondin12,silverman12a}.  
At pre-maximum phases, this supernova (SN) displayed strong absorption features of \ion{Fe}{3} at optical wavelengths, whereas the 
\ion{Si}{2}, \ion{Ca}{2}, and \ion{S}{2} features that typify normal SNe~Ia\footnote{In this paper, a ``normal SN~Ia''
refers to the ``core-normal'' definition of \citet{branch06}.} spectra at these epochs were extremely weak or missing 
altogether.  By two weeks after maximum light, however, the spectrum of SN~1991T was essentially that of a normal Type~Ia event.  The 
decline rate of the $B$-band light curve was measured to be \dm~$= 0.95 \pm 0.05$ mag\footnote{\dm~is defined as the amount in magnitudes that 
the SN fades in the first 15 days following the epoch of $B$-band maximum \citep{phillips93}.} \citep{lira98}, and the absolute magnitude, $M(B)$,
at maximum may have been as much as 0.5~mag brighter than those of normal SNe~Ia with similar decline rates \citep{mazzali95,saha01}.  Like other slow-declining SNe~Ia, the 91T-like
SNe are associated with star-forming galaxies \citep[e.g, see][]{hamuy00}.

A few years after the discovery of SN~1991T, \citet{nugent95} demonstrated that the luminosity versus decline rate relationship for
SNe~Ia was accompanied by a parallel sequence of smoothly varying spectral features at maximum light.  These spectral
variations were ascribed to differences in temperature corresponding to the amount of $^{56}$Ni produced in the explosion,
with 91T-like events representing the highest temperatures (largest $^{56}$Ni masses).  The discovery of SN~1999aa 
\citep{filippenko99}, which showed a pre-maximum spectral evolution intermediate between that of SN~1991T and normal SNe~Ia
\citep{li01b,garavini04}, lent credence to the idea that 91T-like events are simply extreme examples of normal SNe~Ia
\citep{branch01,garavini04}.

\begin{figure*}[t]
\epsscale{.95}
\plotone{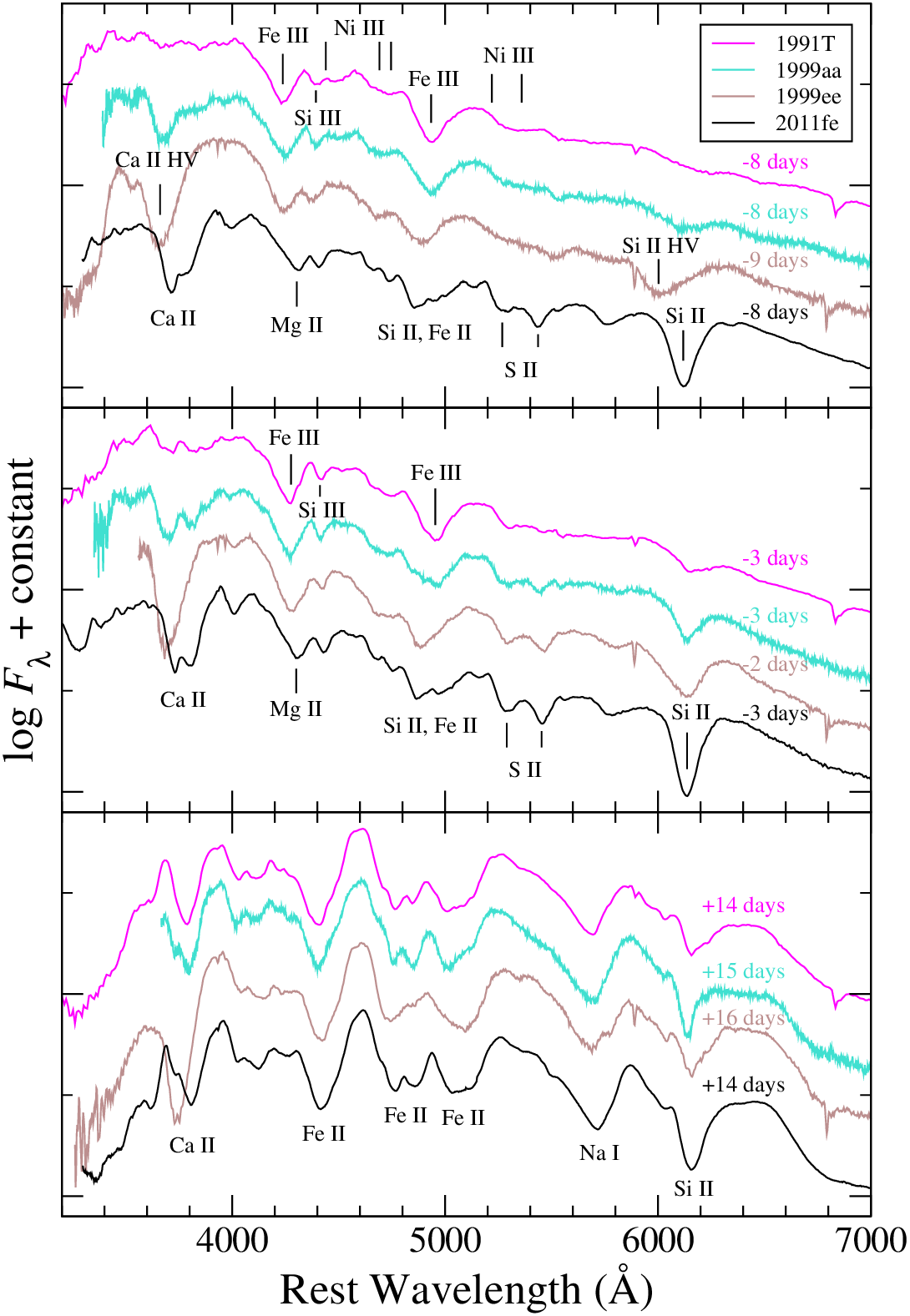} 
\caption{Optical spectra of SN~1991T obtained at $-8$~days, $-3$~days, and $+14$~days with respect to the time of maximum light in the 
$B$-band.  Spectra of SN~1999aa, SN~1999ee, and SN~2011fe at similar phases are plotted for comparison.  The spectra for SN~1991T are taken 
from \citet{phillips92}, those of SN~1999aa from \citet{blondin12}, those of SN~1999ee from \citet{hamuy02b}, and those of SN~2011fe
from \citet{pereira13}.  Each spectrum was corrected for Galactic reddening using the the \citet{schlafly11} recalibration of the \citet{schlegel98} 
infrared dust maps.  In addition, based on \texttt{SNooPy} \citep{burns11} fits to the photometry using the color versus decline rate calibration of \citet{burns14}, 
corrections for host galaxy reddenings of $E(B-V)_{host} = 0.15$ and $0.31$~mag were applied to the spectra of SN~1991T and SN~1999ee,
respectively, while $E(B-V)_{host}$ was assumed to be zero for both SN~1999aa and SN~2011fe.}
\label{fig:91T_99aa_99ee}
\end{figure*}

Figure~\ref{fig:91T_99aa_99ee} displays spectra at three different phases of SN~1991T, SN~1999aa, and 
the normal SNe~1999ee and 2011fe \citep{hamuy02b,stritzinger02}.  At pre-maximum epochs, the spectrum of SN~1991T is remarkable for
the extreme weakness of the \ion{Ca}{2} H~\&~K and \ion{Si}{2}~$\lambda6355$ absorption features, which are clearly visible in the spectra
of SN~1999ee and SN~2011fe, although in the case of the day~$-9$ spectrum of SN~1999ee, the \ion{Ca}{2} and \ion{Si}{2} absorptions 
are dominated by a high-velocity component \citep{mazzali05}.  At these pre-maximum phases, the \ion{Ca}{2} and \ion{Si}{2} absorptions 
in the pre-maxima spectra of SN~1999aa are seen to be intermediate in strength between SN~1991T and SN~1999ee.  At two weeks past
maximum, however, the spectra of all four SNe show only minor differences.  Note that the \dm~values reported in the literature for 
SN~1999aa \citep[$0.85 \pm 0.08$;][]{jha06} and SN~1999ee \citep[$0.94 \pm 0.06$;][]{stritzinger02} are similar to within the errors
to that of SN~1991T, while SN~2011fe had a slightly faster decline rate of $1.10 \pm 0.04$ \citep{pereira13}.

Because they are among the most luminous SNe~Ia, 91T-like events will inevitably be included in distant samples used to study cosmology.  
A better understanding of these objects is therefore crucial not only to deciphering the progenitors and explosion mechanism, but also to avoid 
possible systematic errors in cosmological samples of SNe~Ia.
In this paper, we examine the absolute magnitudes of a sample of 91T-like SNe observed during the course of the second phase of the Carnegie Supernova 
Project \citep[CSP-II;][]{phillips19}. 
In \S\ref{sec:classification}, a new method for classifying them based on 
photometry and one or more optical spectra taken between $-10$ and $+10$~days with respect to the epoch of $B$ maximum is presented.
In \S\ref{sec:CSP-II_sample}, this technique is employed to identify ten 91T-like SNe observed by CSP-II. 
 A brief description of the optical spectral evolution of 91T-like SNe is also included in this section.
Next, in \S\ref{sec:absmags}, the optical and near-infrared (NIR) Hubble diagram residuals of the 91T-like events in the CSP-II sample are compared to 1999aa-like  (henceforth ``99aa-like'') and normal SNe~Ia with similar decline rates.  
Finally, in \S\ref{sec:conclusions}, our conclusions are
summarized. 

\section{Identifying 91T-like Supernovae}
\label{sec:classification}

\begin{figure*}[t]
\epsscale{1.}
\plottwo{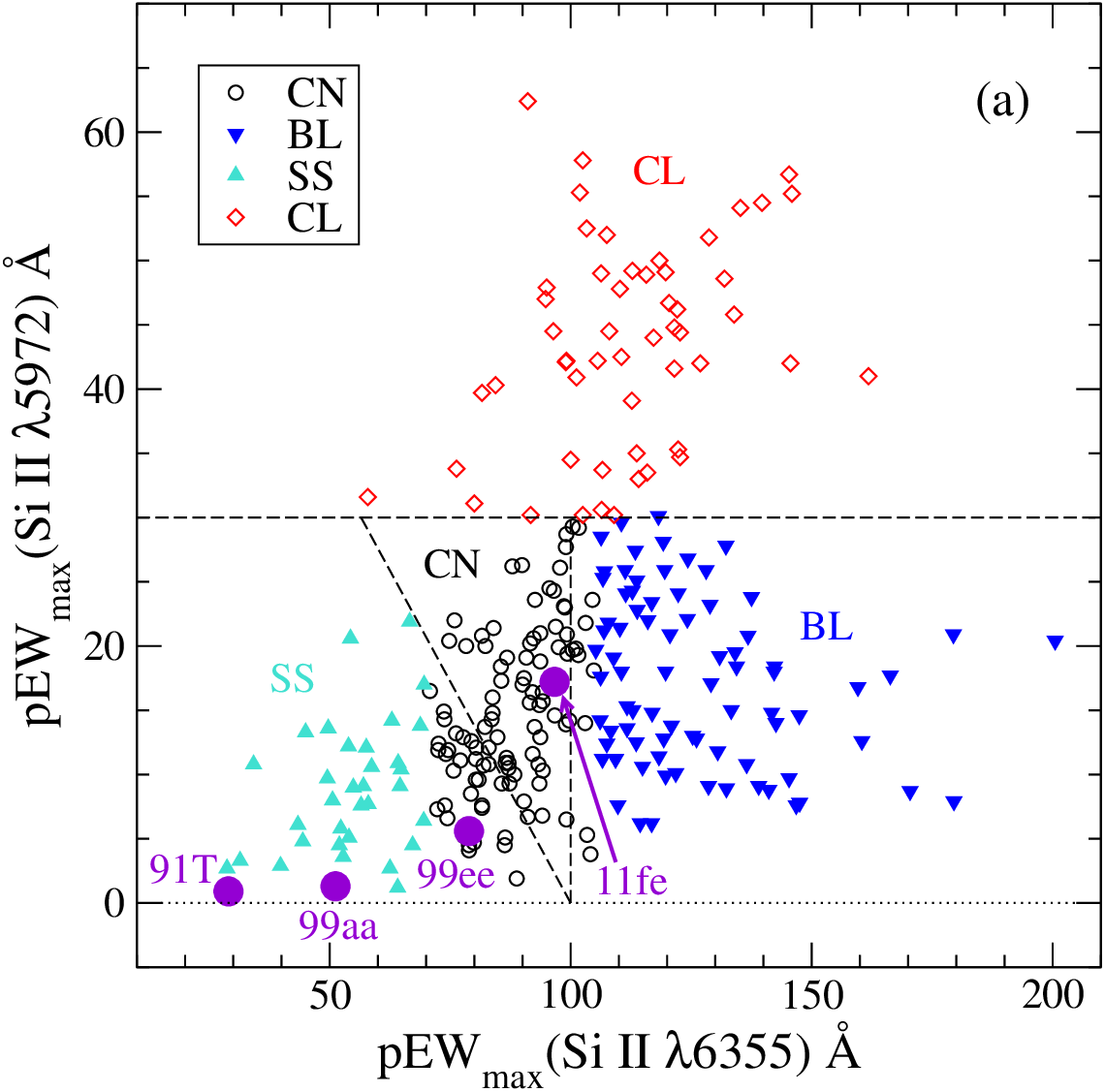}{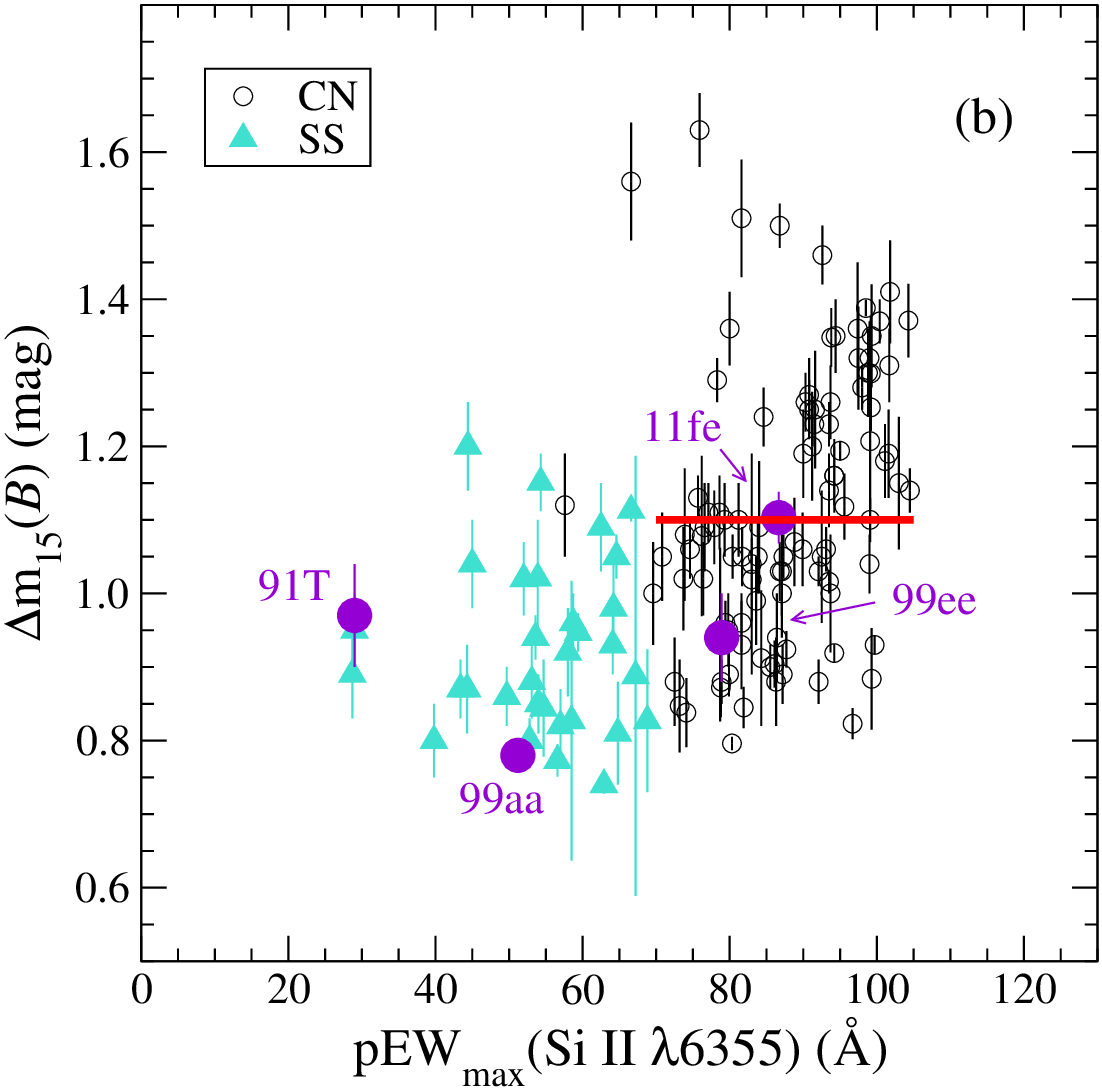}
\caption{(a) Branch diagram for SNe~Ia observed by the CfA Supernova Group and the CSP-I.  The dashed lines approximate the 
boundaries between the different subtypes adopted by \citet{branch09} \citep[See Figure~10 of][]{silverman12b}), while the different colors 
correspond to the boundaries assumed by \citet{folatelli13}, which are similar to those used by \citet{blondin12}; (b) The light curve
decline rate parameter, \dm, plotted versus the pseudo-equivalent width of the \ion{Si}{2}~$\lambda6355$ absorption for the Branch SS 
(green triangles) and CN (black circles) subtypes in the CfA Supernova Group and the CSP-I samples.
In this paper, we refer to CN events lying below the red horizontal line (\dm~$< 1.1$) as the ``slow-declining Branch CN'' SNe (see \S\ref{sec:absmags}).}
\label{fig:branch}
\end{figure*}

\citet{branch06} devised a system for classifying SNe~Ia into four subtypes based on the pseudo-equivalent widths (pEW) of the
\ion{Si}{2}~$\lambda\lambda5972,6355$ absorption features near maximum light.
Figure~\ref{fig:branch}a shows this ``Branch diagram'' 
for SNe~Ia observed spectroscopically by the CfA Supernova Group \citep{blondin12} and the first phase of the  Carnegie Supernova 
Project \citep[CSP-I;][]{folatelli13}.  \citet{branch09} divided the diagram into four areas delineated approximately by the dashed lines 
drawn in Figure~\ref{fig:branch}a: CN for ``core normal'', SS for ``shallow silicon'', BL for ``broad line'', and CL for ``cool''. 
\citet{blondin12} and \citet{folatelli13} divided up the four areas in a slightly different manner; in this paper, we adopt the 
boundaries of \citeauthor{folatelli13} which are indicated by the colors of the symbols in Figure~\ref{fig:branch}a. 
A recent cluster analysis of the Branch diagram by \citet{burrow20} using Gaussian mixture models concluded that the general division into the 
CN, SS, BL, and CL groups was robust and that, when \ion{Si}{2} velocity information is included in the analysis, the BL group appears to be
distinct with respect to the other three groups. 

The positions of SN~1991T and SN~1999aa are shown in Figure~\ref{fig:branch}a.  Both  
fall clearly in the SS portion of the Branch diagram, with SN~1991T having one of the smallest values of pEW(\ion{Si}{2}~$\lambda6355$) of any of the 
SNe plotted.  The positions of the CN SNe~1999ee and 2011fe in the diagram are also shown for reference.

In Figure~\ref{fig:branch}b, the decline rate parameter, \dm, is plotted versus pEW(\ion{Si}{2}~$\lambda6355$) for the SS and CN SNe in the CfA and CSP~I
samples.  The measurements for the CfA sample are taken from \citet{blondin12}, while those for the CSP-I are from \citet{folatelli13}.
Points corresponding to SN~1991T, SN~1999aa, SN~1999ee, and SN~2011fe are indicated.
Not surprisingly, this diagram resembles the region of the Branch diagram occupied by the SS and CN SNe since  
pEW(\ion{Si}{2}~$\lambda5972$) is known to correlate strongly with \dm~\citep{hachinger08,folatelli13}.  This plot clearly
demonstrates that 91T-like SNe, and SS events in general, cannot be distinguished from many CN events on the basis of their decline rates alone.

\begin{figure*}[t]
\epsscale{1.1}
\plotone{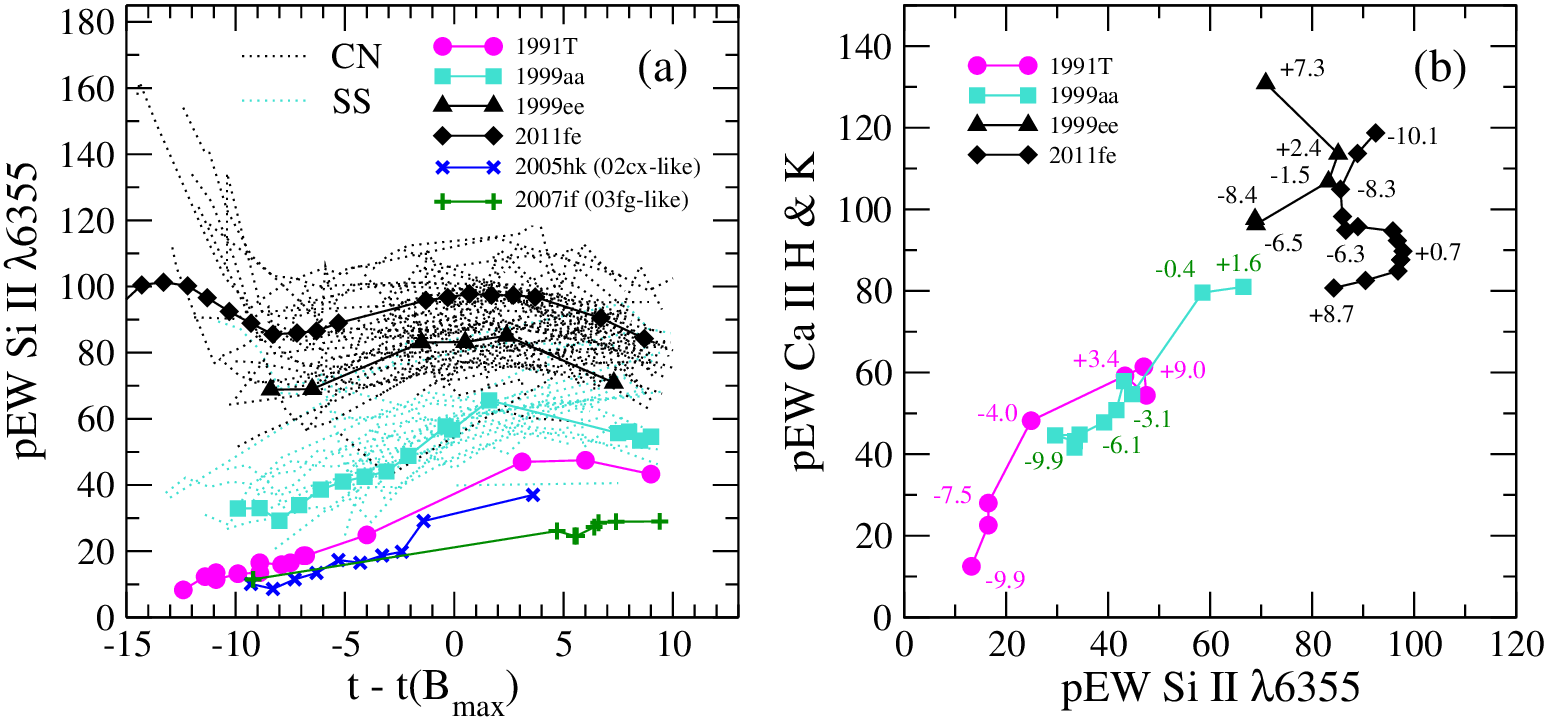}
\caption{(a) Evolution of the pEW(\ion{Si}{2}~$\lambda6355$) for SS (green dotted lines) and CN (black dotted lines)
SNe from the \citet{blondin12} and \citet{folatelli13} spectroscopic samples.  Measurements for SN~1991T, SN~1999aa,
SN~1999ee, and SN~2011fe are highlighted. The spectra for SN~1991T are from \citet{phillips92} and \citet{mazzali95}, those of 
SN~1999aa are from \cite{garavini04} and \citet{blondin12}, those of SN~1999ee are from \citet{hamuy02b} and \citet{blondin12}, and 
those of SN~2011fe are from \citet{pereira13}.  Also included are measurements for the 02cx-like SN2005hk, and for the 03fg-like 
event SN2007if.  The spectra for SN2005hk are from \citet{phillips07} and \citet{blondin12}, and those for SN2007if are from \citet{scalzo10} 
and \citet{blondin12}.; (b) Evolution of the pseudo equivalent widths of the \ion{Ca}{2}~H~\&~K and \ion{Si}{2}~$\lambda6355$ absorption
features as a function of  light curve phase with respect to $t(B_{max})$ for SN~1991T, SN~1999aa, SN~1999ee, and SN~2011fe.
Each point in the diagram represents a spectrum, with a selection of the points labelled with the light curve phase.}
\label{fig:pEW7__pEWCaII_evolution1}
\end{figure*}

The Branch diagram is an excellent tool for identifying 91T-like SNe, but it requires that an optical spectrum be obtained
within a few days of maximum light.  Often this is not the case, and so we have developed an alternative method based on the evolution
of pEW(\ion{Si}{2}~$\lambda6355$).   Figure~\ref{fig:pEW7__pEWCaII_evolution1}a shows our measurements of 
pEW(\ion{Si}{2}~$\lambda6355$) plotted as a function of  phase with respect to the epoch of $B$ maximum, $t-t(B_{max})$, for the CN and SS 
SNe in the CfA and CSP-I spectroscopic samples.  Again, the measurements for SN~1991T, SN~1999aa, SN~1999ee, and 
SN~2011fe are highlighted for reference. For the SS SNe, pEW(\ion{Si}{2}~$\lambda6355$) is observed to increase approximately 
monotonically between $-15$~d~$\la t-t(B_{max}) \la +3$~d, and then stays roughly constant to $+10$~d.  The trajectory of SN~1991T lies 
at the extreme lower edge of the distribution of SS events, while that of SN~1999aa sits more or less in the middle of the SS SNe.  
The evolution of pEW(\ion{Si}{2}~$\lambda6355$) for the CN objects contrasts with that of the SS subtype in that events caught 
early enough display a rapid initial decline in equivalent width until $t-t(B_{max}) \sim -8$~d, followed by a period of slowly increasing values that 
extends to $t-t(B_{max}) \sim +3$~d, and then a final slow decline to $t-t(B_{max}) = +10$~d.  Inspection of the CN spectra obtained 
during the initial sharp decline in pEW(\ion{Si}{2}~$\lambda6355$) indicates that this phenomenon is due to the presence of high-velocity
\ion{Si}{2} absorption at the earliest epochs, which rapidly decays in strength.
Note that this behavior is not observed for the SS SNe, implying either an absence of high-velocity Si and Ca 
 or higher ionization conditions in the outermost ejecta.

In principle, the pseudo equivalent width of the \ion{Ca}{2}~H~\&~K absorption feature can also be used to identify 91T-like SNe.
This is illustrated in Figure~\ref{fig:pEW7__pEWCaII_evolution1}b where pEW(\ion{Ca}{2}~H~\&~K) is plotted versus 
pEW(\ion{Si}{2}~$\lambda6355$) for light curve epochs spanning the range from $-10$~d $\leq t-t(B_{max}) \leq +10$~d
for SN~1991T, SN~1999aa, SN~1999ee, and SN~2011fe.  The phases of selected spectra are labelled to indicate how the 
pseudo equivalent width measurements evolve as a function of time.  The trajectories of the points for SN~1991T and SN~1999aa 
are seen to be similar but offset, with SN~1991T always displaying weaker \ion{Si}{2}~$\lambda6355$ {\em and} \ion{Ca}{2}~H~\&~K 
absorption when compared with spectra at the same phases as SN~1999aa.  This effect is clearly seen in Figure~\ref{fig:91T_99aa_99ee}
where the spectrum of SN~1999aa at $-8$~days is seen to be virtually identical to the spectrum of SN~1991T at $-3$~days.
The trajectories of the CN SNe 1999ee and 2011fe are 
considerably more complex and different -- not just with respect to SN~1991T and SN~1999aa, but also one from the other.  Some of 
these differences are attributable to the presence of high-velocity \ion{Ca}{2} at pre-maximum phases, which is particularly strong in 
SN~1999ee \citep[see Figure~\ref{fig:91T_99aa_99ee} and][]{mazzali05}.

The pseudo equivalent widths for the \ion{Si}{2}~$\lambda6355$ and \ion{Ca}{2}~H~\&~K features were measured
by defining a straight ``continuum'' level between the two flux peaks to either side of the absorption as illustrated in Figure~4 of \citet{folatelli13}.
Errors estimated using the \texttt{IRAF\footnote{IRAF is distributed by the National Optical Astronomy Observatory, which is operated by the 
Association of Universities for Research in Astronomy (AURA) under cooperative agreement with the National Science Foundation.} splot} 
task were typical 20--30\% for the weakest features and 5\% or less for stronger lines.
However, equivalent width measurements are susceptible to additional unaccounted errors in the subtraction of host galaxy light 
from the SN spectrum.  This effect likely explains some of the ``jaggedness'' in the temporal evolution of the 
pEW(\ion{Si}{2}~$\lambda6355$) 
measurements of the CN and SS SNe in Figure~\ref{fig:pEW7__pEWCaII_evolution1}a.  These
errors can make it difficult in some cases to correctly separate 91T-like SNe from 99aa-like events.

In this paper, we use Figure~\ref{fig:pEW7__pEWCaII_evolution1}a to identify true 91T-like SNe.  However, certain peculiar SNe~Ia can masquerade 
as 91T-like events if photometric information is not also available. In particular, the pre-maximum spectra of 02cx-like SNe \citep[also known as 
``Type Iax'' events;][]{foley13} are likewise dominated by \ion{Fe}{3} absorption lines, but at much lower expansion velocities than normal SNe~Ia \citep{li03}.
Figure~\ref{fig:pEW7__pEWCaII_evolution1} shows that the time evolution of the pEW(\ion{Si}{2}~$\lambda6355$) measurements for the well-observed
02cx-like event, SN2005hk, closely mimic those of SN1991T.  However, the 02cx-like events are sub-luminous compared to 91T-like SNe
and, as opposed to 91T-like and normal SNe~Ia, their $i$/$I$-band and NIR light curves peak after $t(B_{max})$ and do not show a 
secondary maximum \citep{li03,phillips07,ashall20}.  Certain 03fg-like SNe\footnote{These SNe are commonly referred to in the literature as 
``Super-Chandrasekhar'' events.  However, since the latter name presumes knowledge of the progenitors which, to date, has not been definitively 
proved, we will refer to them as ``03fg-like'' SNe, after the first recognized example \citep{howell06}.}
can also display 91T-like spectra at pre-maximum phases, the best example being the
well-studied SN2007if \citep{scalzo10,yuan10}.  The evolution of pEW(\ion{Si}{2}~$\lambda6355$) for SN2007if is also plotted in
Figure~\ref{fig:pEW7__pEWCaII_evolution1}a.  Again, such objects can be discriminated from 91T-like SNe from their exceptionally broad $i$/$I$-band and 
NIR light curves that peak after $t(B_{max})$ and lack of a clear secondary maximum \citep{ashall20,ashall21}.

To summarize, we employ both spectroscopic and photometric requirements for classifying a SN as a 91T-like event, which we define
as follows:

\begin{itemize}
 
  \item Spectroscopic:  At least one pEW(\ion{Si}{2}~$\lambda6355$) measurement before 10~days after $B$ maximum that is consistent with the trajectory of SN~1991T
   in Figure~\ref{fig:pEW7__pEWCaII_evolution1}a.  In practical terms, this translates to pEW(\ion{Si}{2}~$\lambda6355$) $<$ $\sim$20~\AA\ at $t-t(B_{max}) = -10$~d,
   $<$ than $\sim$40~\AA\ at maximum, and $<$ $\sim$50~\AA\ at $t-t(B_{max}) = +10$~d.
   
   \item Photometric:  An $i$/$I$-band light curve that reaches maximum {\em before} the epoch of $B$ maximum and displays a clear secondary maximum.
   
\end{itemize}

Before ending this section, it is important to emphasize that programs  providing automated spectral classification such as SNID \citep{blondin07} make little
distinction between 91T-like and 99aa-like SNe.  Indeed, the standard SNID template subgroups are ``Ia-norm'', ``Ia-91T'', ``Ia-91bg'', ``Ia-csm'', and ``Ia-pec''.
Hence, many SNe that SNID classifies as ``Ia-91T'' are actually ``99aa-like'' events.

\section{91T-like Supernovae in the CSP-II Sample}
\label{sec:CSP-II_sample}

Using the above-described criteria, we have identified ten 91T-like events observed by the CSP-II.
The CSP-II was carried out between 2011~October and 2015~May, during which high-quality optical and 
NIR light curves were obtained for 214 SNe~Ia, 125 of which were located in the smooth Hubble flow at 
redshifts $0.027 < z < 0.137$ \citep{phillips19}.  These data are being published separately in the final 
data release of CSP-II photometry \citep{suntzeff22}.
Details concerning these ten 91T-like SNe are given in Appendix~\ref{sec:csp}, including plots of the temporal 
evolution of pEW(\ion{Si}{2}~$\lambda6355$) and the individual light curves.

\begin{figure*}[t]
\epsscale{1.1}
\plotone{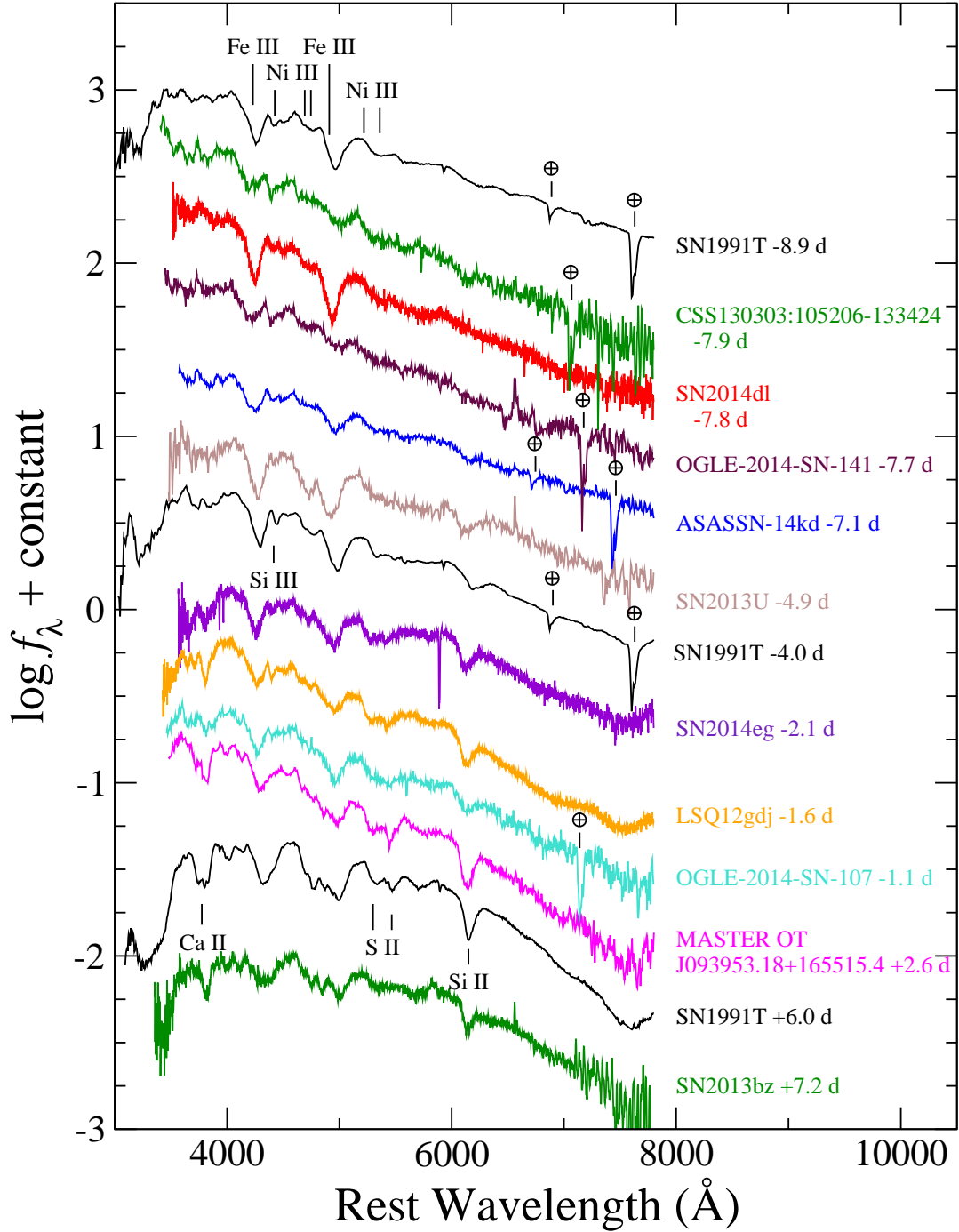}
\caption{Optical spectra of the ten 91T-like SNe in the CSP-II sample obtained $-8$ to $+7$~days with respect to the time of maximum light in the 
$B$-band.  Spectra of SN1991T at three phases covering this range are plotted in black for comparison.  Where necessary, the spectra have been adjusted via a 
low-order polynomial to match the photometry. They have also been corrected for Galactic reddening using the \citet{schlafly11} recalibration of the 
\citet{schlegel98} infrared-based dust map and assuming $R_V = 3.1$.  Unremoved telluric absorption features are marked with an $\oplus$ symbol.}
\label{fig:CSPII_91T}
\end{figure*}

Selected optical spectra of the 91T-like SNe identified in the CSP-II sample are found in Figure~\ref{fig:CSPII_91T}.  
This figure illustrates that the spectra of events obtained at similar phases are quite similar overall.  The largest differences are in the slopes (i.e., the colors)
of the spectra, which are consistent with differences in host galaxy reddening.
At pre-maximum epochs, the spectra are dominated
by the \ion{Fe}{3}~$\lambda$4404 and $\lambda$5129 absorption features.  Note, however, that the overall strengths of the \ion{Fe}{3} lines can vary considerably from object to object.
A prime example of this is illustrated in Figure~\ref{fig:CSPII_91T}, where the spectrum of SN2014dl is observed
to have very strong \ion{Fe}{3} absorption at $-7.8$~days, while these same lines are nearly washed out in the spectra of CSS130303:105206-133424 at
a phase of $-7.9$~days and OGLE-2014-SN-141 at $-7.7$~days.  In between these two extremes is the spectrum of ASASSN-14kd obtained at $-7.1$~days.
As noted by \citet{sasdelli14}, the \ion{Fe}{3} absorption observed at pre-maximum epochs in 91T-like SNe must be produced mostly by stable $^{54}$Fe
since $^{56}$Co would not have had enough time since explosion to decay to $^{56}$Fe.  These authors concluded that the amount of stable Fe required to produce the observed \ion{Fe}{3}
absorption in SN~1991T is consistent with solar metalicity material in the progenitor.
It is tempting to speculate that the observed variations in the strength of the \ion{Fe}{3} absorption at pre-maximum epochs could reflect
differences in the metallicities of the progenitors.  It seems unlikely that this is an ionization effect since the strength of the \ion{Si}{3}~$\lambda$4564 absorption (see below) shows less 
variation than \ion{Fe}{3}~$\lambda$4404 in spite of the very similar ionization potentials of Fe$^{++}$ and Si$^{++}$.
And while differing amounts of host galaxy light contamination in the spectra might produce such variations, there is no evidence to support this, 
particularly in the case of CSS130303:105206-133424 which exploded in the outer region of its faint host.

Other weaker absorption features visible between and redward of the \ion{Fe}{3}~$\lambda\lambda$4404,5129 lines at these early phases 
(see top panel of Figure~\ref{fig:91T_99aa_99ee}) have been ascribed to transitions of \ion{Ni}{3} \citep{ruiz-lapuente92,mazzali95,fisher99,hatano02}.  
The $-8.9$~days spectrum of SN~1991T plotted in  Figure~\ref{fig:CSPII_91T} shows a strong absorption feature at 3200~\AA\ which, 
according to \citet{sasdelli14}, is a blend of \ion{Co}{3} and \ion{Fe}{3} lines, with the Co arising from the decay of $^{56}$Ni.  These authors concluded that 
Ni, Co, and Fe must be present to velocities as high as $\sim$17,000~\kms\ to match the shape of the spectral features produced by these elements.

Absorption due to multiplet~2 of \ion{Si}{3} ($\lambda_0 \simeq 4564$~\AA) is present in the pre-maximum spectra of 91T-like events beginning at the earliest epochs
\citep{jeffery92,mazzali95}.
However, \ion{Si}{2}~$\lambda$6355 absorption does not become clearly visible until approximately a week before maximum light, and \ion{Ca}{2}~H~\&~K and the iconic ``W''
absorption feature of \ion{S}{2} at $\sim$5400~\AA\ are undetectable until only a few days before maximum light.  The presence of \ion{Si}{3} absorption early on
demonstrates that intermediate-mass elements are present in the outermost layers of 91T-like SNe, but are largely invisible due to the unusually high radiation temperatures at these phases
\citep{mazzali95,nugent95,sasdelli14}. This may also account for the lack of observable high-velocity \ion{Ca}{2} and \ion{Si}{2}
in the outer ejecta of 99aa-like and 91T-like SNe.

As mentioned in \S\ref{sec:intro} and illustrated in the bottom panel of Figure~\ref{fig:91T_99aa_99ee}, by two weeks after maximum, the optical spectra of 91T-like SNe 
begin to closely resemble those of normal SNe~Ia.  
P-Cygni lines of \ion{Ca}{2}~H~\&~K and the $\lambda\lambda$8498,8542,8662 triplet are clearly visible, and \ion{Fe}{2} 
features begin to dominate much of the spectrum \citep[e.g., see Figure~10 of][]{jeffery92}.  Broad absorption is present at $\sim$7500~\AA\ \citep[see Figure~5 of][]{filippenko92}
which several authors have identified with \ion{O}{1}~$\lambda$7774 \citep{filippenko92,jeffery92,sasdelli14}.  Unfortunately, at these phases, the strength of the \ion{Si}{2} absorption is no longer
a useful tool for separating 91T-like events from 99aa-like and slow-declining Branch CN SNe.

\section{Absolute Magnitudes}
\label{sec:absmags}

As highlighted in the introduction to this paper, and also summarized by \cite{taubenberger17} in his review article, 91T-like SNe have long been 
suspected of being intrinsically brighter than ``normal'' SNe~Ia at the luminous end of the absolute magnitude versus decline rate relation.  In the
case of SN~1991T itself, if we assume the Cepheid-based distance modulus of $30.74 \pm 0.12$ (statistical) $\pm 0.12$ (systematic) measured
by \citet{saha01} and use the \texttt{SNooPy EBV\_model2} \citep{burns11} to fit the $UBVRI$ photometry published by \citet{lira98},
a host reddening of $E(B-V) = 0.15 \pm 0.01$~mag and 
an absolute magnitude of $M_V = -19.62 \pm 0.22$
are implied.  This is $0.23 \pm 0.23$~mag brighter than a SN with \sbv~$=1.1$\footnote{The color stretch parameter, \sbv, is a dimensionless parameter defined as the time difference between $B$-band 
maximum and the reddest point in the $(B-V)$ color curve divided by 30 days, where typical SNe~Ia have \sbv $\sim$1 \citep{burns14}.}
according to the $M_V$ versus \sbv\ relation for SNe~Ia given
by \citet{burns18} for events with \sbv$> 0.5$ and $E(B-V)_{host} < 0.5$~mag.  While interesting, the large error in this measurement does not
allow a definitive conclusion.  A  better approach would be to compare a sample of 91T-like SNe in the smooth Hubble flow 
where the redshift can be used as a precise relative measurement of distance.  The CSP-II sample of SNe~Ia provides
such an opportunity.

\begin{figure}[h]
\epsscale{0.7}
\plotone{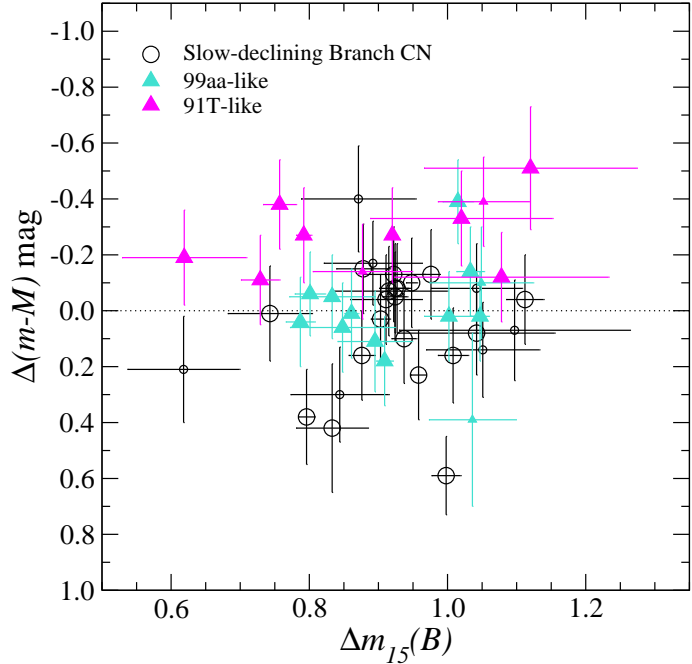}
\caption{Hubble diagram distance modulus residuals, $\Delta(m-M)$, for 91T-like, 99aa-like, and 
slow-declining Branch CN SNe observed by the CSP-II.  
The sample of slow-declining Branch CN SNe is defined as those events
with \dm~$< 1.10$.  The larger symbols are
SNe whose values of \dm\ were measured directly; the smaller symbols correspond to those where \texttt{SNooPy} template fits were used to
measure the decline rate.  The samples were limited to those with redshifts $z \geq 0.02$ to minimize uncertainties due to peculiar velocities.
Note that negative values of $\Delta(m-M)$ correspond to overluminous events.}
\label{fig:Hubble_resid_dm15}
\end{figure}

In Figure~\ref{fig:Hubble_resid_dm15}, we plot Hubble diagram residuals, $\Delta(m-M)$, versus \dm\ for the 91T-like and 99aa-like SNe observed by the CSP-II.
The Hubble diagram residual is defined as the difference between the distance modulus predicted from the host galaxy redshift 
using a best-fit cosmological model\footnote{Assumes standard $\Lambda$CDM cosmology and a fixed Hubble constant $H_0 = 72~{\rm km~s^{-1}~Mpc^{-1}}$, density parameter $\Omega_m = 0.27$, and cosmological constant
parameter $\Omega_\Lambda = 0.73$.} to the full set of SNe~Ia observed by the CSP-I and CSP-II, and the distance modulus derived from the intrinsic
color analysis described in detail by \citet{burns18}.  
For comparison, the Hubble diagram residuals for the slow-declining Branch CN SNe in the CSP-II sample, which
we define as those events with \dm~$< 1.10$, are also plotted in Figure~\ref{fig:Hubble_resid_dm15}.  
Negative values therefore indicate more luminous SNe Ia.  This figure
 clearly shows that all of the 91T-like SNe have negative Hubble diagram residuals, implying that they are over-luminous by
$\sim0.1$--$0.5$~mag.  It also confirms that 91T-like (and 99aa-like) SNe cannot be distinguished from slow-declining Branch CN events with similar 
decline rates on the basis of their \dm\ measurements alone.

\begin{figure}[h]
\epsscale{0.8}
\plotone{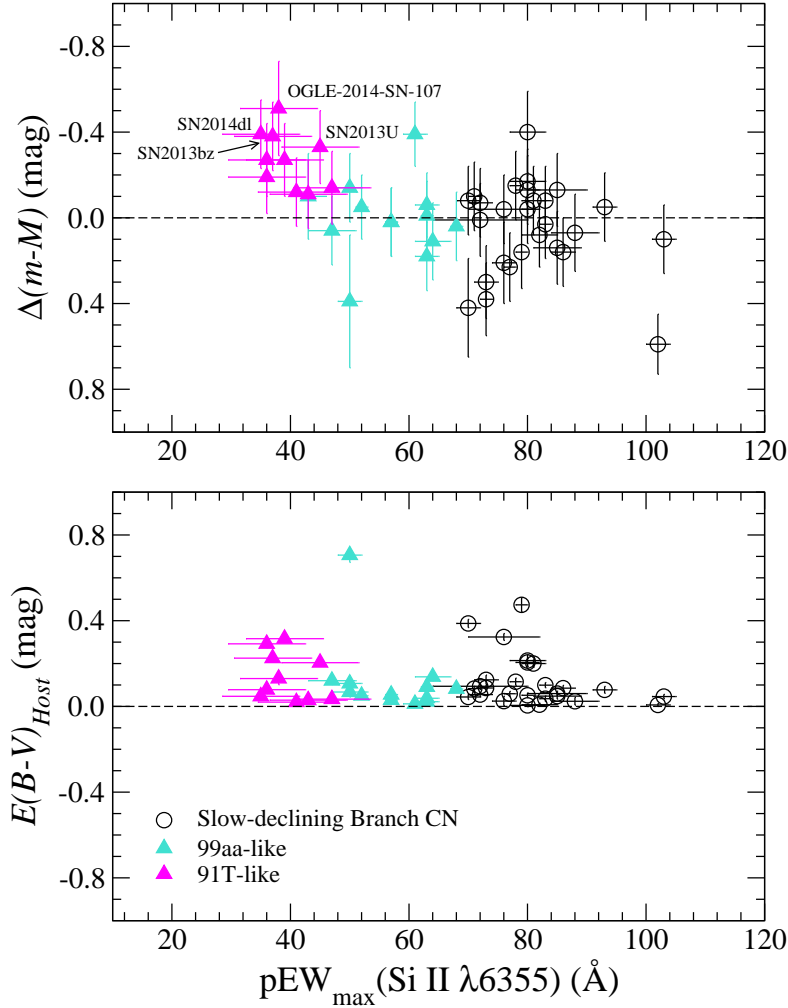}
\caption{(above) Hubble diagram residuals in magnitudes plotted versus the pseudo equivalent width of the \ion{Si}{2}~$\lambda6355$ absorption
at maximum light for 91T-like, 99aa-like, and CN SNe observed by the CSP-II.
(below) $E(B-V)_{host}$ values derived using \texttt{SNooPy} for the same SNe.}
\label{fig:Hubble_resid}
\end{figure}

To separate the 91T-like events from the 99aa-like and slow-declining Branch CN SNe,  Hubble diagram residuals are plotted versus pEW(\ion{Si}{2}~$\lambda6355$) 
at maximum light in Figure~\ref{fig:Hubble_resid}.   The pEW(\ion{Si}{2}~$\lambda6355$) values for the 99aa-like and slow-declining Branch CN SNe 
are taken from \citet{morrell22}, while for the 91T-like SNe, a low-order polynomial fit to the shape of the evolution of the equivalent width 
values for SN~1991T was used to extrapolate the pEW  measurements to maximum light.  The error bars reflect the average error obtained 
in applying the same extrapolation method to the individual measurements of pEW(\ion{Si}{2}~$\lambda6355$) for SN~1991T.
Again, we see that the distance modulus residuals of the 99aa-like and Branch CN SNe scatter about zero, 
whereas the values for the 91T-like SNe are systematically negative. 

As will be presented in a future paper, the $(B-v)$ and $(r-i)$ colors of the 91T-like SNe at maximum light appear to be similar to those 
of the 99aa-like and slow-declining Branch CN SNe.
Nevertheless, the possibility must be considered that the apparent overluminous nature of the 91T-like events is the result of
overestimates of the effect of host galaxy reddening due to
incorrect assumptions of the colors of these extreme SNe.  In the lower half of Figure~\ref{fig:Hubble_resid} we plot the
$E(B-V)_{host}$ values returned by \texttt{SNooPy} for the  SNe plotted in the upper half of this figure.  The mean host 
reddening for the 91T-like SNe is $\sim$0.06~mag higher than the average values for the 99aa-like and slow-declining Branch CN events 
(see Table~\ref{tab:mean_params}).  Two of the four most luminous 91T-like objects, SN~2013U and SN~2013bz, have relatively large
values of $E(B-V)_{host}$ ($0.20 \pm 0.02$ and $0.23 \pm 0.02$~mag, respectively).  However, the two others, OGLE-2014-SN-107 and SN~2014dl, have more modest values of $E(B-V)_{host} = 0.13 \pm 0.03$~mag and $0.05 \pm 0.01$~mag. 

\begin{figure}[h]
\epsscale{.95}
\plotone{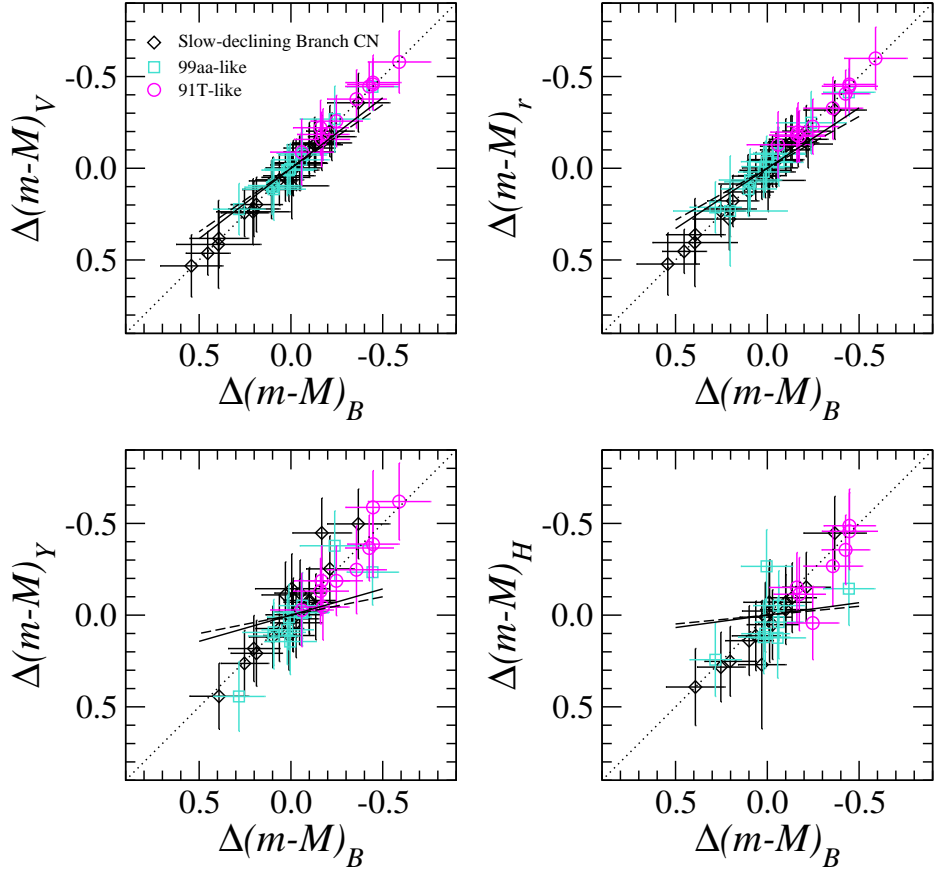} 
\caption{$B$-band Hubble diagram residuals in magnitudes plotted versus the residuals in $V$, $r$, $Y$, and $H$ for 91T-like, 99aa-like, and 
and CN SNe observed by the CSP-II.  The solid black line in each plot indicates the expected correlation in the
residuals for $R_V = 3$ and a systematic error between $\pm 0.5$~mag in $E(B-V)$; the dashed black line shows the same for $R_V = 2$.}
\label{fig:Hubble_resid2}
\end{figure}

A way of checking whether errant host galaxy dust corrections are responsible for the systematically-higher luminosities of the 91T-like
SNe compared to the 99aa-like and slow-declining Branch CN events is to look at the Hubble diagram residuals as a function of wavelength.  In particular, 
we would expect dust absorption to have a much smaller effect on the Hubble diagram residuals for the NIR filters compared to the optical.
The values of $\Delta(m-M)$ plotted in the upper half of Figure~\ref{fig:Hubble_resid} are derived from combining the \texttt{SNooPy} fits to the $uBgVriYJH$
light curves.  However, we can also look at the residuals as a function of the individual filters.  Table~\ref{tab:mean_params} indicates that
the weighted mean Hubble diagram residuals for the 91T-like, 99aa-like, and CN samples are remarkably constant from the optical to the NIR.
This is displayed graphically in Figure~\ref{fig:Hubble_resid2} where the residuals in $V$, $r$, $Y$, and $H$ for each SN are plotted versus the
residuals in $B$.  Excellent correlations are observed in all four bands, with the 91T-like SNe consistently occupying the most-luminous part.

\begin{deluxetable}{lrcrcrc}
\tabletypesize{\scriptsize}
\tablecolumns{7}
\tablewidth{0pt}
\tablecaption{Mean values of  \dm, $E(B-V)_{host}$, and $\Delta(m-M)$ for CSP-II SNe~Ia \label{tab:mean_params}}
\tablehead{
\colhead{Parameter} &
\colhead{91T-like} &
\colhead{n} &
\colhead{SS} &
\colhead{n} &
\colhead{CN} &
\colhead{n} 
}
\startdata
\dm & 0.87 (0.17) & 10 & 0.92 (0.11) & 13 &  0.92 (0.10) & 27 \\
$E(B-V)_{host}$ & 0.12 (0.11) & 10 & 0.07 (0.19) & 13 & 0.06 (0.13) & 27 \\
\hline
$\Delta(m-M)_B$ & $-0.31$ (0.18) & 10 & $-0.01$ (0.19) & 13 & $+0.04$ (0.22) & 27 \\
$\Delta(m-M)_V$ & $-0.33$ (0.16) & 10 & $-0.03$ (0.19) & 13 & $+0.05$ (0.22) & 27 \\
$\Delta(m-M)_r$ & $-0.32$ (0.16) & 10 & $-0.03$ (0.18) & 13 & $+0.05$ (0.21) & 27 \\
$\Delta(m-M)_Y$ & $-0.27$ (0.21) & 10 & $+0.02$ (0.21) & 11 & $-0.01$ (0.22) & 20 \\
$\Delta(m-M)_H$ & $-0.25$ (0.19) & 7 & $+0.02$ (0.17) & 8 & $+0.03$ (0.21) & 15 \\
\hline
$\Delta(m-M)_{uBgVriYJH}$\tablenotemark{a} & $-0.33$ (0.18) & 10 & $-0.01$ (0.15) & 13 & $+0.05$ (0.22) & 27 \\
\enddata
\tablecomments{All values shown are weighted means. The root mean square dispersion is given in parentheses, followed by
the number of SNe, ``n'', used in calculating the mean.}
\tablenotetext{a}{Average cross-validated distance modulus calculated from the $u$, $B$, $g$, $V$, $r$, $i$, $Y$, $J$, and $H$ filters
\citep[see][]{burns18}}
\end{deluxetable}

The \texttt{SNooPy color\_model} method used to calculate the Hubble diagram 
residuals in Table~\ref{tab:mean_params} and Figures~\ref{fig:Hubble_resid_dm15}--\ref{fig:Hubble_resid2}
employs intrinsic colors from \citet{burns18} to constrain $E(B-V)$ and
$R_V$ for each SN.  These values, in turn, are used to compute reddening-corrected
magnitudes.  Errors in $E(B-V)$ and $R_V$ will therefore propagate a
systematic error in the Hubble diagram residual in all filters. However, such
errors should grow smaller with wavelength (see Figure~\ref{fig:Hubble_resid2}) and cannot account for the
strong, slope~$=$~1, correlations observed in Figure~\ref{fig:Hubble_resid2} which are dominated by the peculiar
velocities of the host galaxies.  Moreover, except for OGLE-2014-SN-141, the \texttt{color\_model} method gives
acceptable fits in the $BVriYJH$ bands (see Appendix~\ref{sec:csp}).
Still, it is important to confirm
that the high luminosities of 91T-like SNe in all filters are not an
artifact of the \texttt{color\_model} method.  The Hubble diagram residuals for CSP-II
SNe~Ia published by \citet{uddin22} provide an alternative check since 
these were derived by applying the empirical \citet{tripp98} luminosity--color
relation to maximum light magnitudes derived from
the \texttt{SNooPy max\_model} method.
This analysis yields average residuals in {\em all} filters of 
approximately $-0.22 \pm 0.07$~mag, $-0.04 \pm 0.05$~mag, and $+0.08 \pm
0.04$~mag for the 91T-like, 99aa-like, and slow-declining Branch CN
SNe, respectively, confirming
that 91T-like SNe are systematically more luminous than their 99aa-like and slow-declining Branch CN counterparts.  
Figures~\ref{fig:Hubble_resid_dm15}--\ref{fig:Hubble_resid2}, as well as Table~\ref{tab:mean_params},
also suggest that 99aa-like SNe may be intermediate in luminosity between 91T-like and slow-declining Branch CN events, but this will require future confirmation 
from larger samples.

\section{Conclusions}
\label{sec:conclusions}

91T-like SNe are the extreme members of the Branch SS group in displaying
the weakest \ion{Si}{2}$~\lambda6355$ and \ion{Ca}{2} H~\&~K absorption
absorption at maximum light of any SNe~Ia.  We show that these objects
can be identified from one or more optical spectra by plotting
pEW(\ion{Si}{2}~$\lambda6355$) versus light curve phase for 
$t-t(B_{max}) \leq +10$~days.  They are  differentiated from 02cx-like
and 03fg-like SNe in possessing $i/I$-band light curves that reach
maximum before the of epoch $B$ maximum, and that also display a clear
secondary maximum.  Lacking accurate knowledge of
light curve phase, 91T-like SNe may be easily confused with the less
extreme 99aa-like events.

From a sample of ten 91T-like SNe observed during the course of the
CSP-II, we find clear evidence that 91T-like SNe are over-luminous by
$\sim$0.1--0.5~mag compared to 99aa-like and slow-declining Branch CN SNe.  This
difference in luminosity is remarkably constant from optical to NIR
wavelengths, arguing that overestimates of the host galaxy dust
corrections cannot explain this finding.  The data further suggest that
99aa-like events may be intermediate in luminosity between 91T-like SNe
and slow-declining Branch CN SNe, although this requires
further confirmation.

Based on an analysis by \citet{leaman11} of 726 SNe discovered by the Lick Observatory Supernova Search (LOSS), 
\citet{li11} concluded that 9\% of all SNe~Ia in the local Universe are 91T-like events.  However, \citeauthor{li11} employed a broad definition
for what constituted a ``Ia 91T'' that included 99aa-like events.  In any event, they are likely to be sufficiently 
rare in the local Universe that their effect on measurements of the Hubble constant should be small.
This is borne out by an analysis of the CSP-II sample using the methodology of \citet{uddin22}.  Restricting the sample to the 178 SNe Ia with
\sbv~$> 0.8$ so as to exclude lower-luminosity events, and then repeating the calculation
after removing the ten 91T-like events, the value of the Hubble constant changes by only 0.1\%.
However, there is no assurance that the frequency of 91T-like remains constant at
greater and greater look-back times.  And even if their relative numbers have not changed as the Universe has evolved, we can 
expect to observe proportionately more of them at the magnitude limits of discovery surveys due to Malmquist bias. Unfortunately,
the light curve width alone cannot be used to identify 91T-like SNe, although this may not be true at ultraviolet wavelengths \citep{jiang18}. 
At least for now, spectroscopy combined with rest frame $i$/$I$-band photometry remains the most reliable tool for discriminating between 91T-like, 99aa-like, and slow-declining Branch CN SNe.

\begin{acknowledgments}
The work of the CSP-II has been generously supported by the National Science Foundation under 
grants AST-1613426, AST-1613455, and AST-1613472.
The CSP-II was also supported in part by the Danish Agency for Science and Technology and Innovation through a 
Sapere Aude Level 2 grant. M. Stritzinger acknowledges  funding by a research 
grant (13261) from VILLUM FONDEN.
E.B. is supported in part by NASA grant 80NSSC20K0538.
L.G. acknowledges financial support from the Spanish Ministerio de Ciencia e Innovaci\'on (MCIN), the Agencia Estatal de Investigaci\'on (AEI) 10.13039/501100011033, and the European Social Fund (ESF) "Investing in your future" under the 2019 Ram\'on y Cajal program RYC2019-027683-I and the PID2020-115253GA-I00 HOSTFLOWS project, from Centro Superior de Investigaciones Cient\'ificas (CSIC) under the PIE project 20215AT016, and the program Unidad de Excelencia Mar\'ia de Maeztu CEX2020-001058-M.
We gratefully acknowledge the use of WISeREP -- https://wiserep.weizmann.ac.il.
M.M.P. thanks Richard Scalzo for interesting discussions of extreme SNe~Ia.
This research has made use of the NASA/IPAC Extragalactic Database (NED), which is funded by the National 
Aeronautics and Space Administration and operated by the California Institute of Technology.
Finally, we thank the referee for suggestions on improving the text and figures.
\end{acknowledgments}

%

\vspace{5mm}
\facilities{Magellan:Baade (IMACS imaging spectrograph, FourStar wide-field near-infrared camera, 
Magellan:Clay (LDSS3 imaging spectrograph), Swope (SITe3 CCD imager, 
e2v 4K x 4K CCD imager), du~Pont (SITe2 CCD imager, Tek5 CCD imager, WFCCD imaging spectrograph, RetroCam 
near-infrared imager), 
NOT (ALFOSC), 
La Silla-QUEST, CRTS, PTF, iPTF, OGLE, ASAS-SN, PS1, KISS, ISSP, MASTER, SMT)}


\software{IRAF\citep{tody86,tody93}, SNooPy \citep{burns11}}



\pagebreak

\appendix

\restartappendixnumbering

\section{91T-like SNe from the CSP-II}
\label{sec:csp}

\begin{figure}
\epsscale{0.9}
\plotone{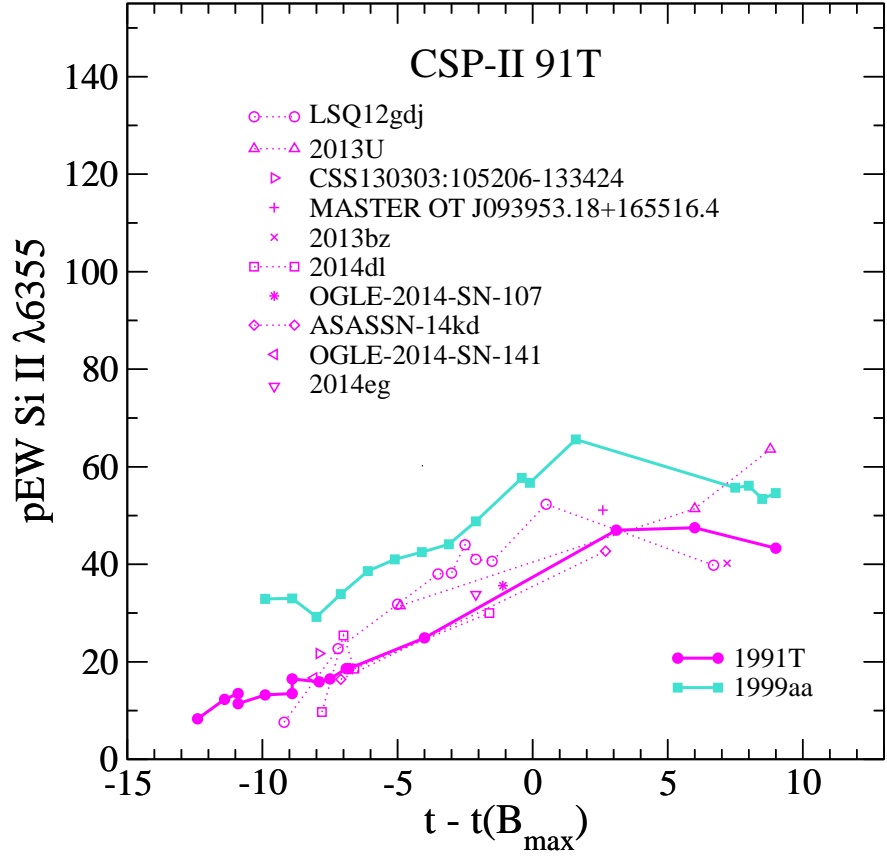}
\caption{Evolution of pEW(\ion{Si}{2}~$\lambda6355$) for 91T-like SNe identified in the CSP-II sample.  
The trajectories of SN~1991T and SN~1999aa in this diagram are plotted for reference.}
\label{fig:pEW7_evolution_CSPII}
\end{figure}

\begin{figure}
\epsscale{1.1}
\plottwo{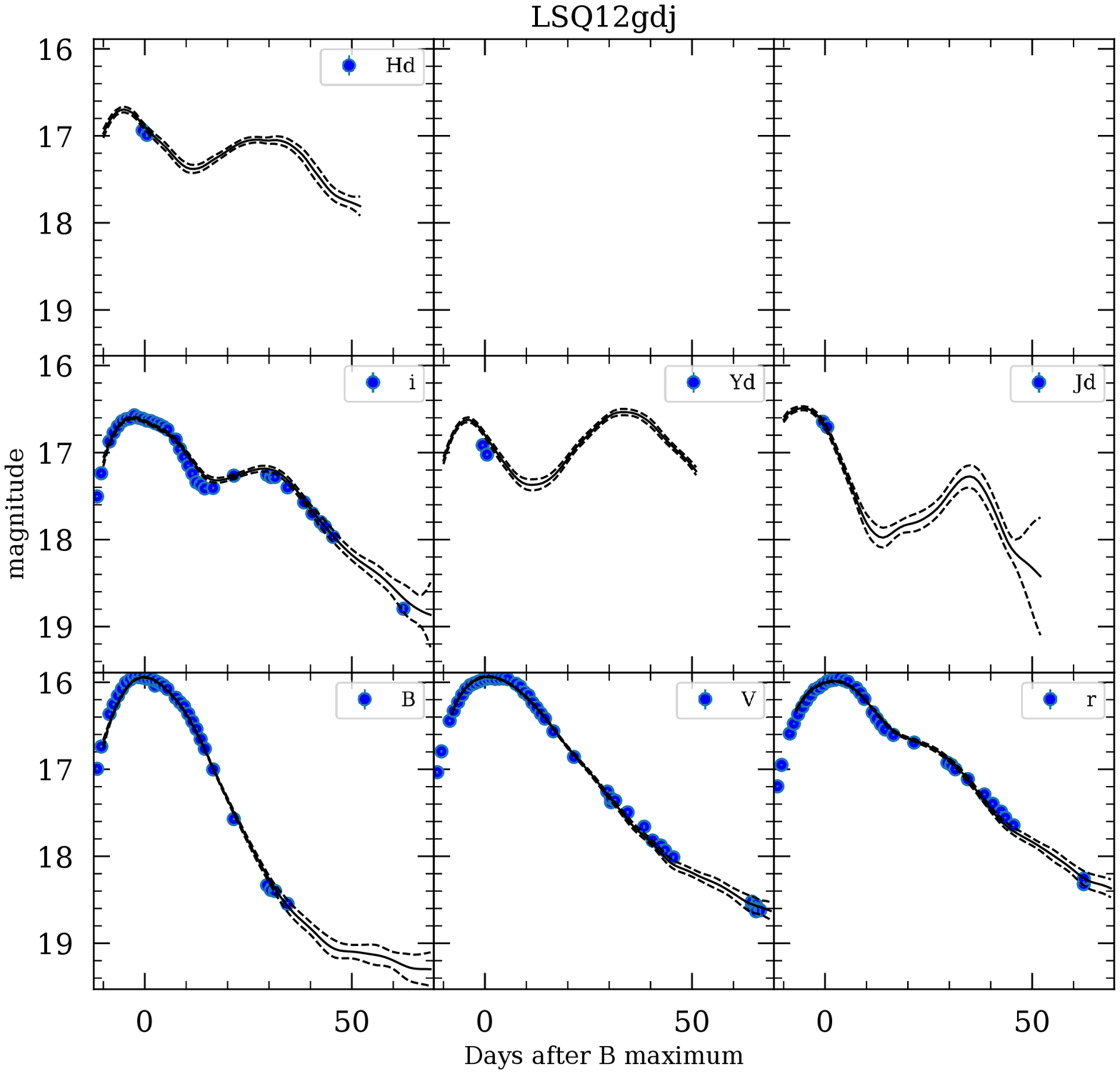}{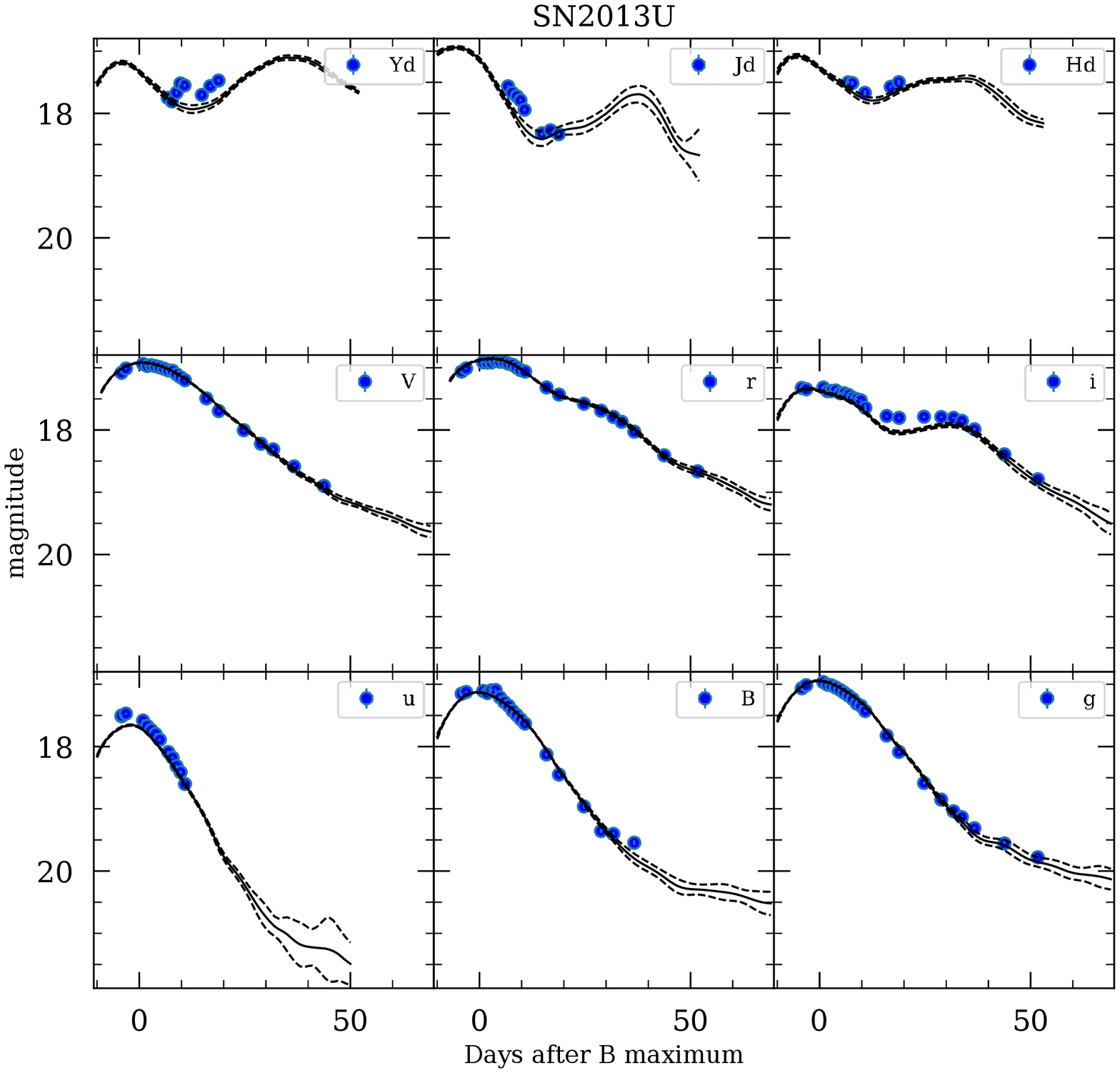}
\plottwo{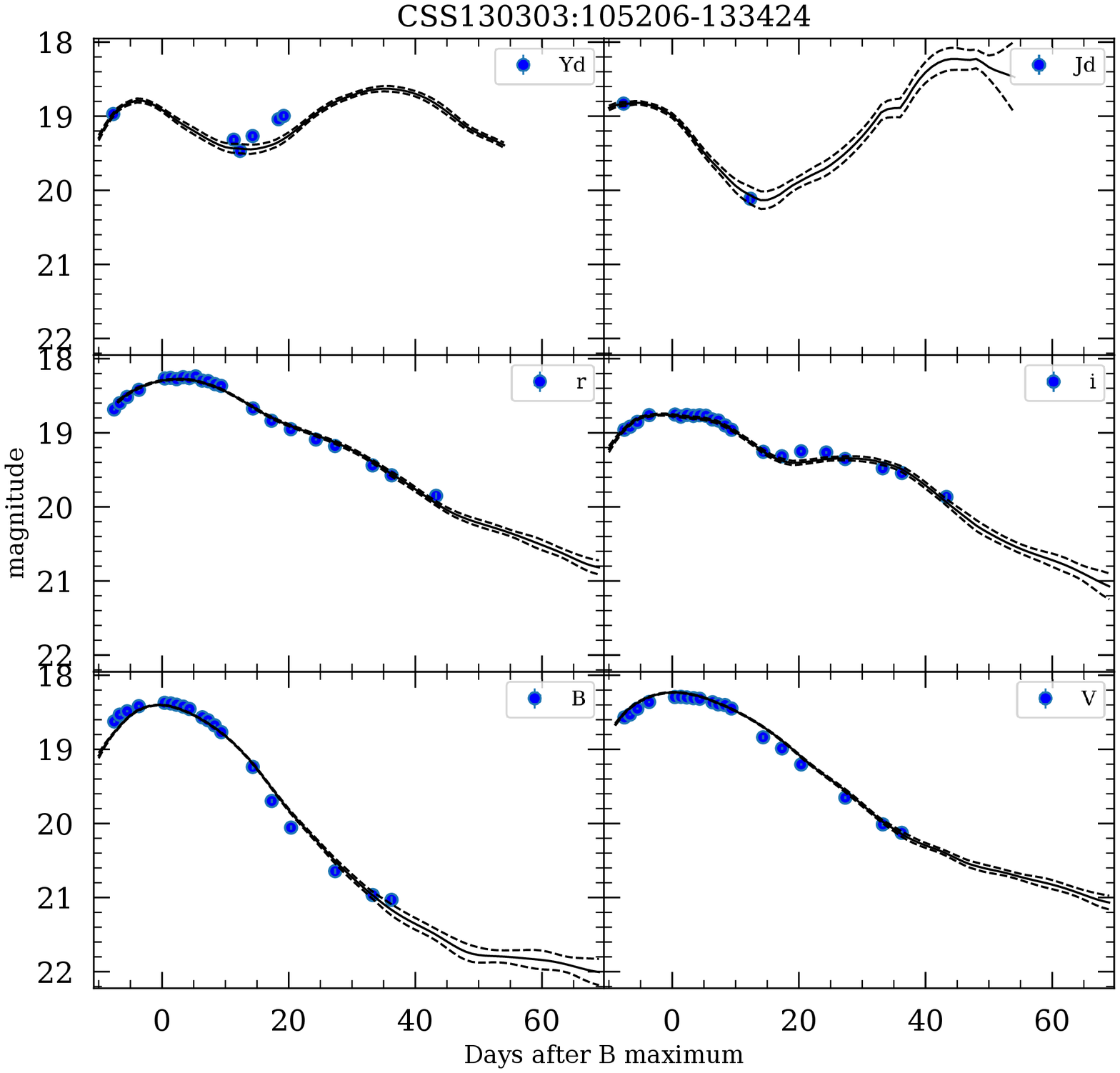}{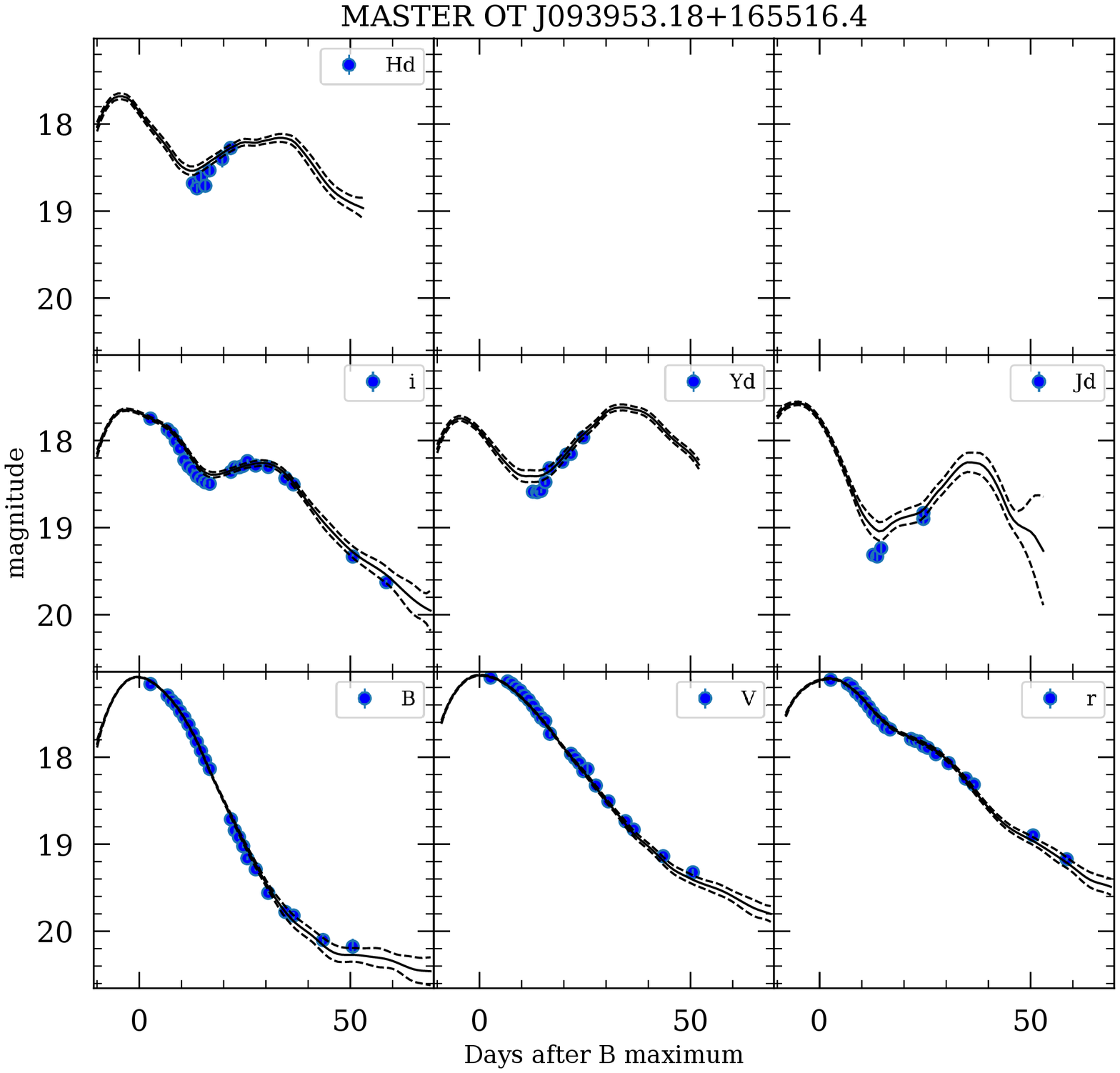}
\caption{SNooPy fits to the 91T-like SNe LSQ12gdj, SN2013U, CSS130303:105206-133424, and SN~2014eg derived using the intrinsic color analysis described by \citet{burns18}.  Magnitudes plotted on the y-axis for each filter are as observed.  The x-axis gives the time in days
{\em in the observer frame} after the epoch of $B$ maximum.}
\label{fig:snpy_plots1}
\end{figure}

\begin{figure}
\epsscale{1.1}
\plottwo{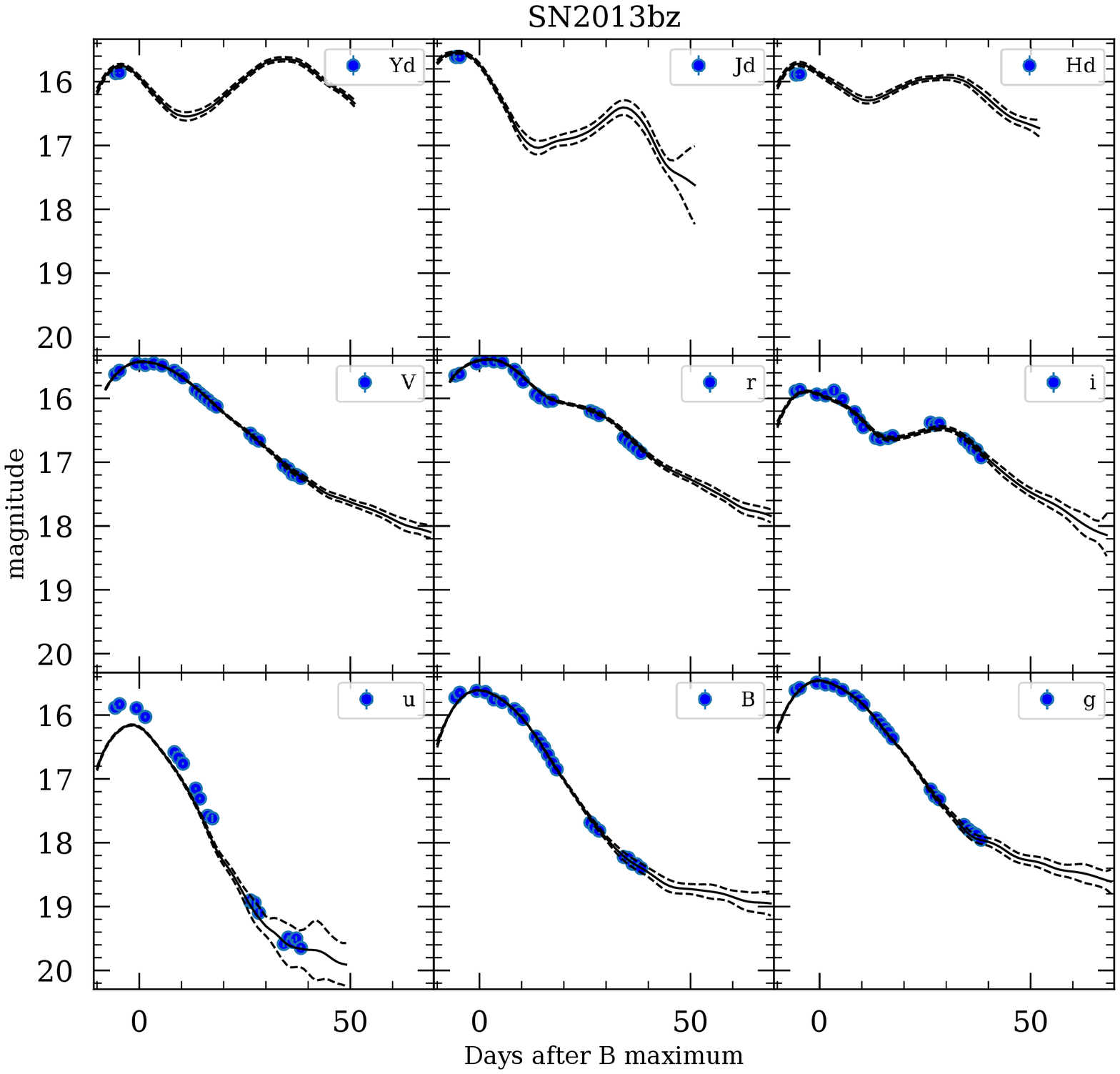}{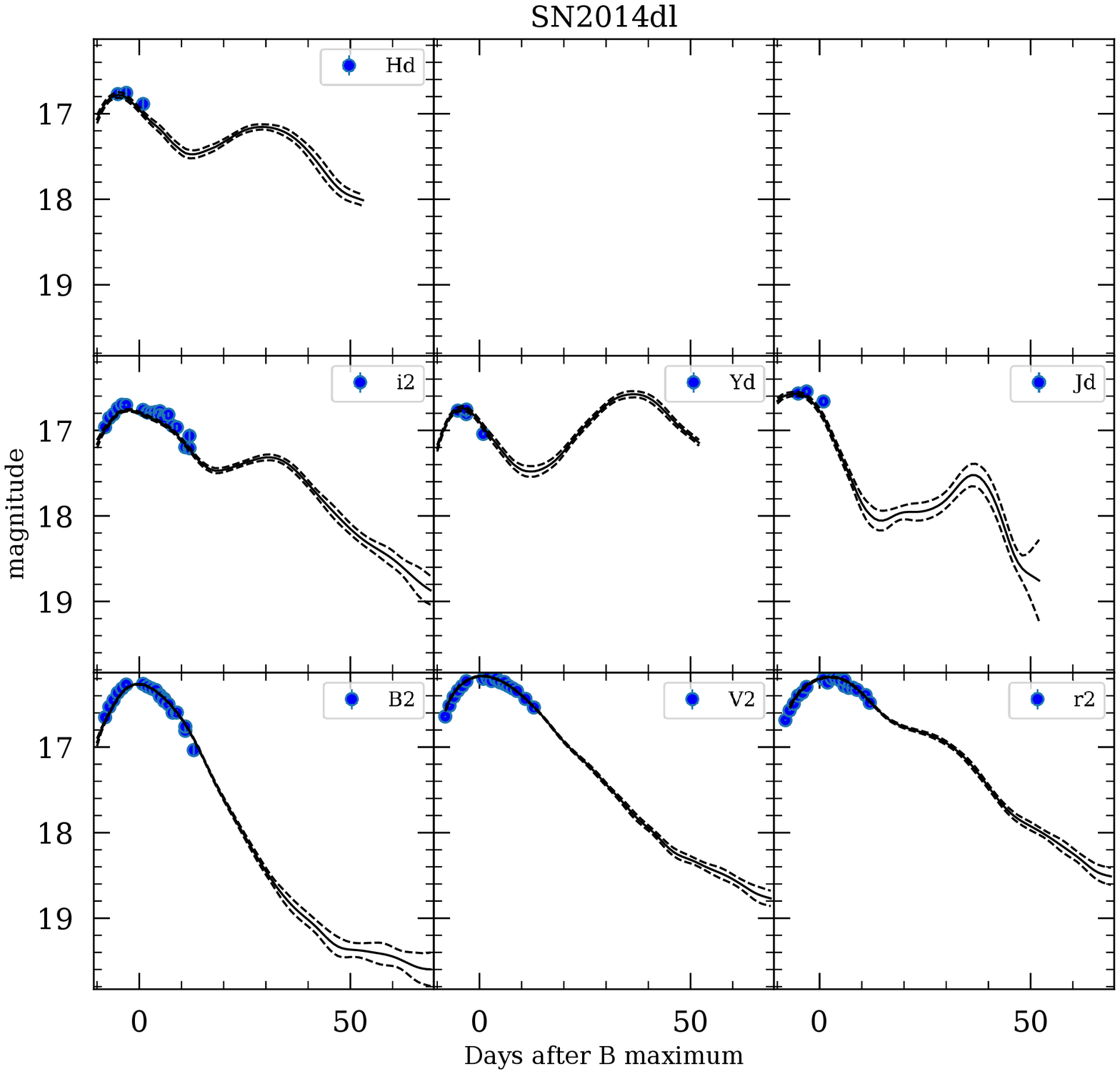}
\plottwo{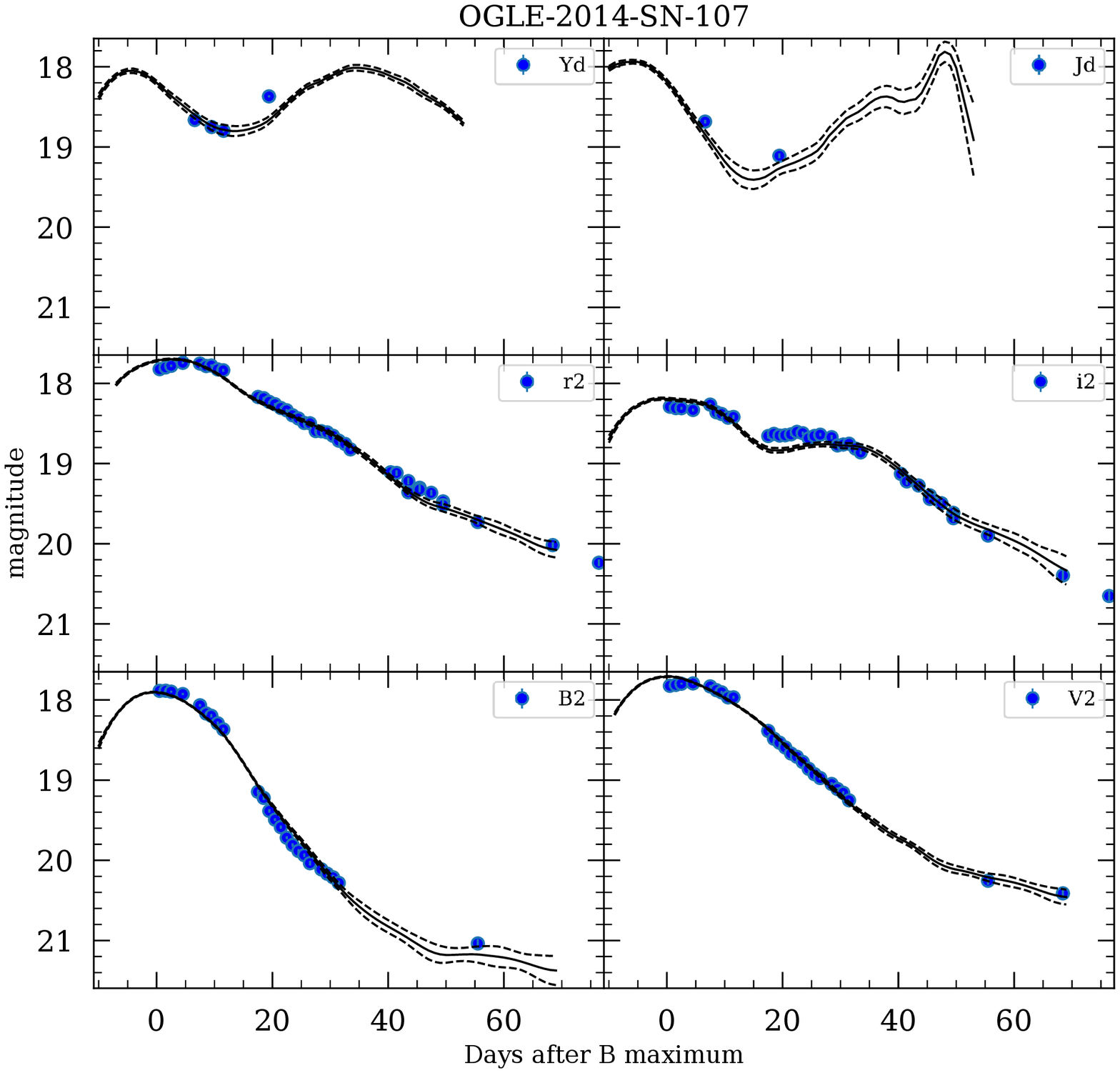}{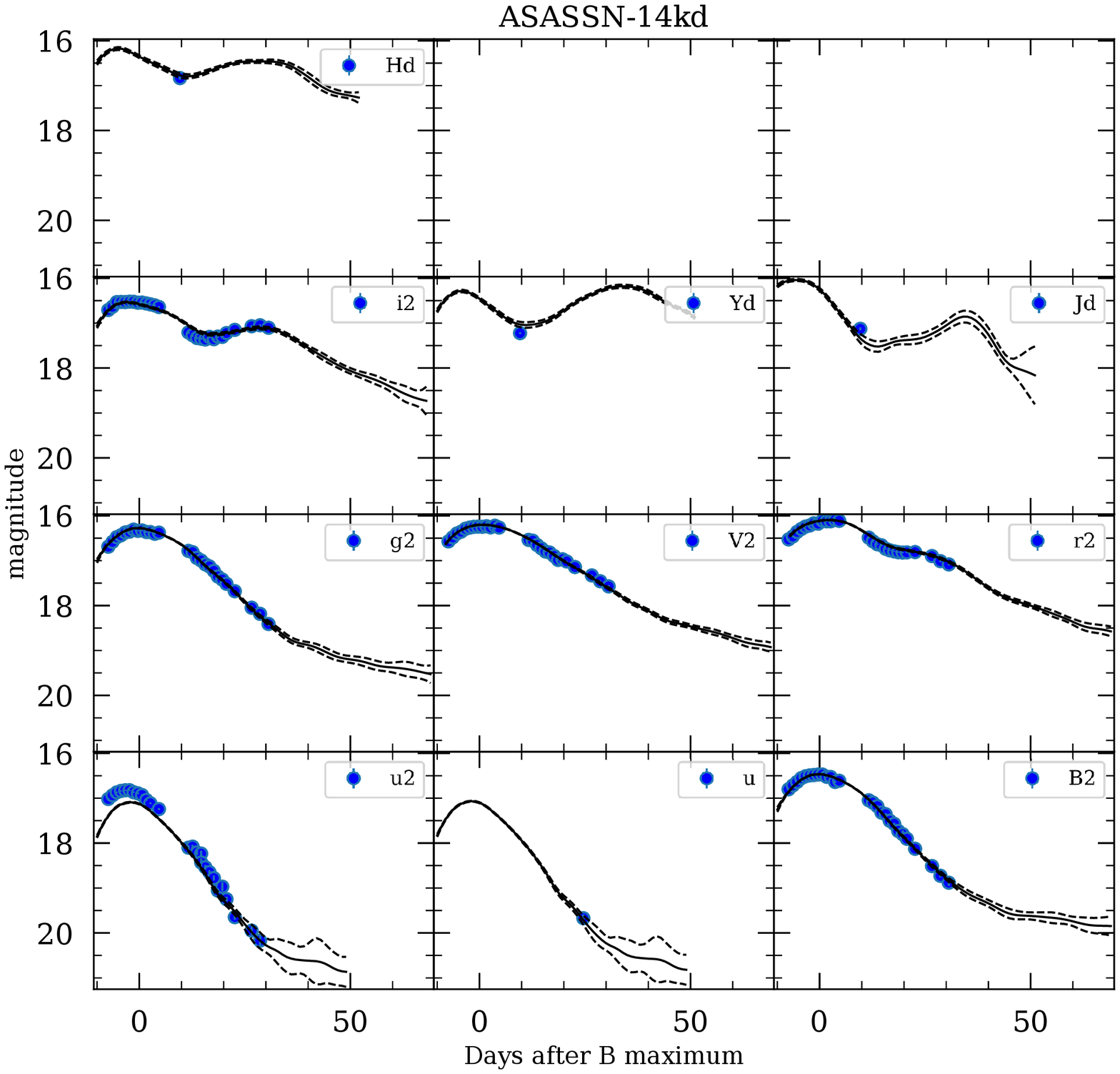}
\caption{Same as Figure~\ref{fig:snpy_plots1}, except for the 91T-like SNe 2013bz, 2014dl, OGLE-2014-SN-107, and ASASSN-14kd.}
\label{fig:snpy_plots2}
\end{figure}

\begin{figure}
\epsscale{1.1}
\plottwo{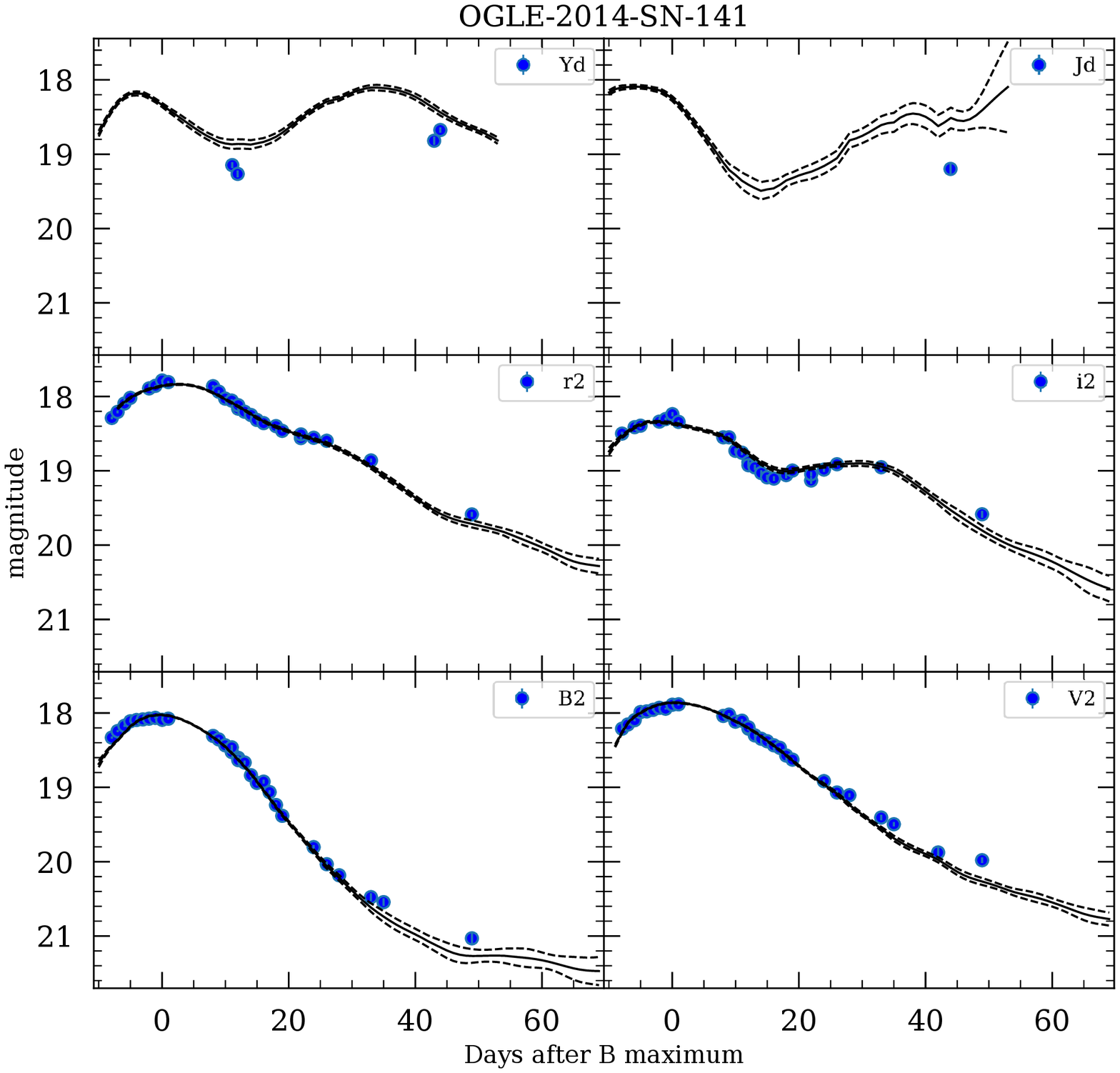}{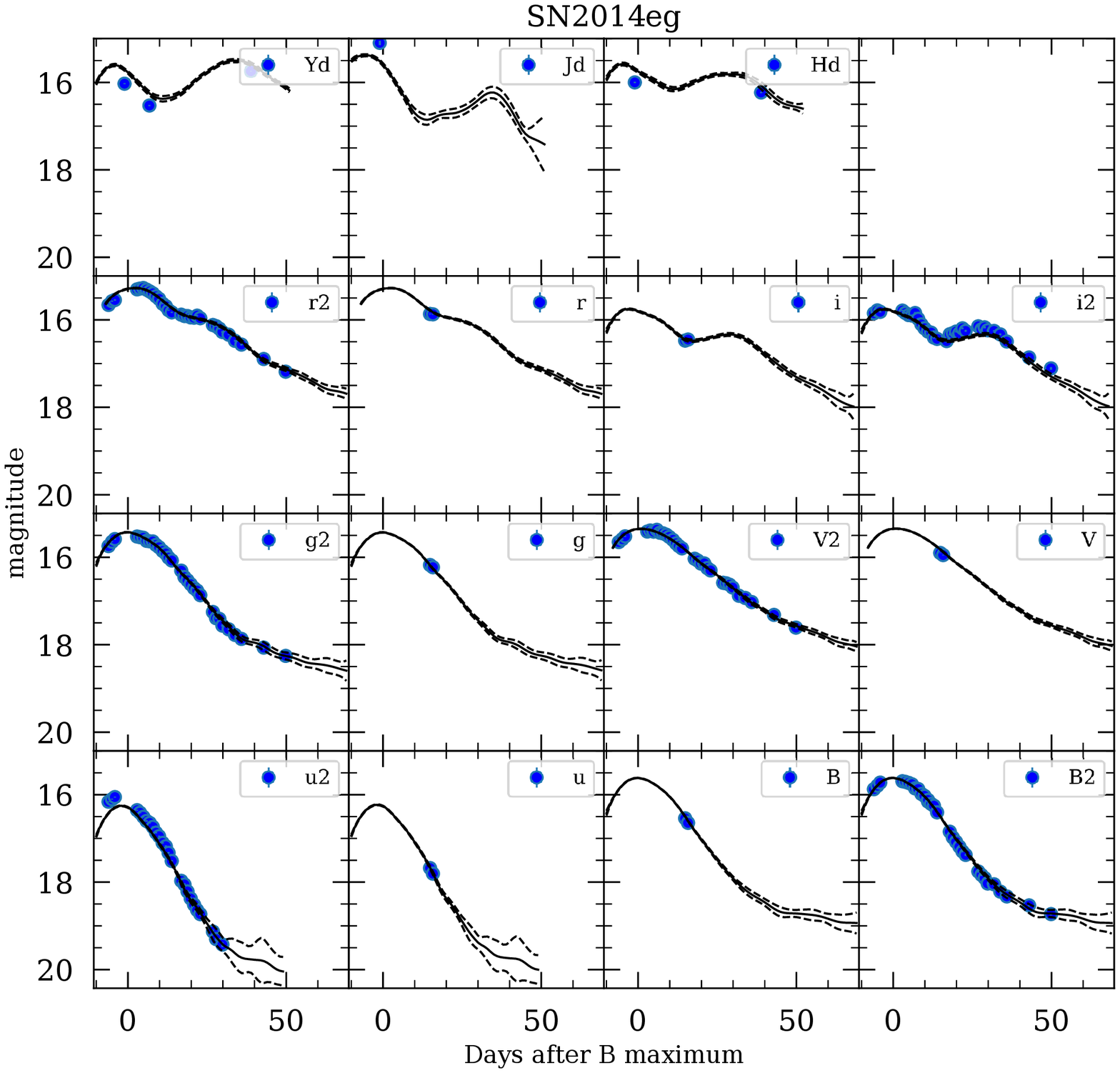}
\caption{Same as Figure~\ref{fig:snpy_plots1}, except for the 91T-like SNe OGLE-2014-SN-141 and SN~2014eg.}
\label{fig:snpy_plots3}
\end{figure}

A total of ten 91T-like events in the CSP-II sample were identified using the  spectroscopic and photometric criteria defined in 
\S\ref{sec:classification}.  These objects are listed in Table~\ref{tab:csp2_91T}.
Figure~\ref{fig:pEW7_evolution_CSPII} shows the temporal 
evolution of pEW(\ion{Si}{2}~$\lambda6355$) for the CSP-II sample, and, in Figures~\ref{fig:snpy_plots1}--\ref{fig:snpy_plots3},
\texttt{SNooPY color\_model} fits are plotted.
These fits employ the intrinsic colors from \citet{burns18} to constrain $E(B-V)$ and
$R_V$ for each SN.

Comments on individual objects are as follows:

\begin{itemize}

\item LSQ12gdj:  Extensive observations of this SN have been previously published by \citet{scalzo14}, who concluded that it was ``spectroscopically 91T-like''.  Figure~\ref{fig:pEW7_evolution_CSPII}
shows that the evolution of the pseudo equivalent width of \ion{Si}{2}~$\lambda6355$ was more similar to SN~1991T than SN~1999aa.  The light curves
plotted in Figure~\ref{fig:snpy_plots2} also clearly show that LSQ12gdj had a strong secondary maximum in the $i$ band, and that primary maximum
in $i$ was reached before $B$~maximum. 

\item SN~2013U: This SN was discovered by the Puckett Observatory Supernova Search\footnote{\url{http://www.cometwatch.com/supernovasearch.html}}
and was correctly classified as a 91T-like event from a spectrum obtained with the Asiago 1.82-m Copernico Telescope \citep{gagliano13,tomasella13}.  The \texttt{SNooPy color\_model} templates displayed in Figure~\ref{fig:snpy_plots2} 
provide excellent fits at maximum light in the $BVgri$ filters, but underpredict the observed brightness in $u$.  Comparing with the other SNe with $u$-band photometry plotted in 
Figure~\ref{fig:snpy_plots2},
this appears to be a general characteristic of 91T-like SNe \citep[see also][]{scalzo14}.

\item CSS130303:105206-133424: This object was discovered by the Catalina Real-Time Transient Survey \citep[CRTS;][]{djorgovski11}.  A spectrum 
obtained by the Public ESO Spectroscopic Survey of Transient Objects \citep[PESSTO;][]{smartt15} with the ESO La Silla 3.6 m NTT was classified by
\citet{Blagorodnova13} as corresponding to a SN~Ia caught $\sim$7~days before maximum light.  As discussed in \S\ref{sec:CSP-II_sample}, this spectrum
is peculiar even for a 91T-like event in showing only very weak absorption features, including those of \ion{Fe}{3}.

\item MASTER OT J093953.18+165516.4: The discovery of this SN by the MASTER network of robotic telescopes \citep{gorbovskoy13} was 
announced by \citet{shumkov13}.  It was independently discovered by the CRTS with the designation MLS130215:093953+165516.  A PESSTO
spectrum obtained by \citet{benitez-herrera13} was stated to show ``similarity with that of the peculiar, 1991T-like SN~1998es'' near maximum
light.  SN~1998es was actually a 99aa-like event according
to the classification method presented in \S\ref{sec:classification}.  However, as seen in Figure~\ref{fig:pEW7_evolution_CSPII}, measurement 
of pEW(\ion{Si}{2}~$\lambda6355$) from this spectrum of MASTER OT J093953.18+165516.4 is consistent with a 91T-like event a few 
days after maximum light.  Technically speaking, the CSP-II $i$-band photometry plotted in Figure~\ref{fig:snpy_plots2} does not begin early enough to confirm that maximum was 
reached before maximum light in $B$, but the excellent fits
to the \texttt{SNooPy} templates leave little doubt that this was a bonafide 91T-like SN.

\item SN~2013bz: This is another SN discovered by the CRTS \citep{howerton13} which designated it as SNhunt188.  A spectrum taken with 
the Asiago 1.82-m Copernico Telescope by \citet{ochner13} was said to be ``a good match ... with the so-called `super-Chandrasekhar'
SNe Ia 2006gz and 2009dc.''  However, a spectrum obtained ~4~days earlier with the 4.2-m William Herschel Telescope by Chen et al. 
\citep{howerton13} was stated to be consistent with either a normal SN~Ia at maximum light, or a 91T-like event a week after maximum.
The CSP-II light curves indicate that this spectrum, which is not publicly available, was obtained at $+3$~days.  The value of 
pEW(\ion{Si}{2}~$\lambda6355$) measured from the Asiago spectrum is fully consistent with a 91T-like classification (see 
Figure~\ref{fig:pEW7_evolution_CSPII}).  This SN has a striking excess at $u$-band maximum (see Figure~\ref{fig:snpy_plots2}).

\item SN~2014dl: Yet another SN discovered by the CRTS \citep{drake14}. Spectra obtained at the Asiago and Las Campanas Observatories were found to be similar to that
of a 91T-like event $\sim$1~week before maximum \citep{drake14}.

\item OGLE-2014-SN-107: This object was discovered in the course of the OGLE-IV Real-Time Transient Search \citep{wyrzykowski14}.
From a PESSTO spectrum, \citet{takats14} classified it as a 91T-like SN observed approximately $\sim$5~days before maximum light. 
This is confirmed in Figure~\ref{fig:pEW7_evolution_CSPII}.  As illustrated in Figure~\ref{fig:snpy_plots2}, the minimum between the primary and secondary maxima of the $i$-band
light curve of this SN is particularly shallow.

\item ASASSN-14kd: The discovery of this SN in the course of the All-Sky Automated Survey for Supernovae (ASASSN) \citep{shappee14,kochanek17} 
was announced by \citep{nicholls14}.  From a PESSTO spectrum obtained within hours of discovery, \citet{firth14} identified it as a SN~Ia that was
best matched by SNID 91T-like templates at a phase of $-7$~days, which we confirm in Figure~\ref{fig:pEW7_evolution_CSPII}.

\item OGLE-2014-SN-141: This is another SN discovered by the OGLE-IV Real-Time Transient Search.  A single PESSTO spectrum was
obtained by \citet{dimitriadis14}, who classified it as a SN~Ia at a phase of approximately $-7$~days and noted that it was an ``excellent
match to a SN1991T-like SN~Ia.''  This is confirmed in Figure~\ref{fig:pEW7_evolution_CSPII}. Interestingly,
although the \texttt{SNooPy color\_model} gives a good fit to the $BVRi$ light curves with a host galaxy color excess of $0.08 \pm 0.02$~mag and $R_V = 3.4 \pm 0.9$,
the $Y$ and $J$ templates over predict the brightness in 
the NIR by $\sim$0.4--0.6~mag.

\item SN2014eg: This SN was serendipitously discovered in ESO 154-G010 by PESSTO observers who were obtaining follow-up observations
of the Type II-n SN~2013fc in the same host galaxy.  A spectrum was obtained six nights later by \citet{smith14} who classified SN2014eg as
a 91T-like caught approximately one week before maximum.  Extensive spectroscopic and photometry observations of this SN may be 
found in \citet{dimitriadis17}. 

\end{itemize}

\begin{deluxetable*}{lllllc}[h]
\tabletypesize{\scriptsize}
\tablecolumns{6}
\tablewidth{0pt}
\tablecaption{91T-like Supernovae Observed by the CSP-II\label{tab:csp2_91T}}
\tablehead{
 \colhead{SN Name} &
\colhead{Host Galaxy} &
\colhead{$z_{\rm{helio}}$\tablenotemark{a}} &
\colhead{\dm \tablenotemark{b}} &
\colhead{\sbv \tablenotemark{c}} &
\colhead{Spectroscopy\tablenotemark{d}}
}
\startdata
LSQ12gdj                                                    &   ESO 472-G 007                                 &   $0.0303$ &   0.73 (0.03)         &  1.14 (0.05)  &  1        \\
SN2013U                                                     &   CGCG 008-023                                   &   $0.0345$ &  1.02 (0.13)         &  1.25 (0.05)   &  2,3,4  \\
CSS130303:105206-133424                       &   GALEXASC J105206.27-133420.2    &   $0.0789$ &  1.08 (0.16)         &  1.19 (0.05)   &  5       \\
MASTER OT J093953.18+165516.4          &   CGCG 092-024                                   &   $0.0478$ &   0.84 (0.04)         &  1.12 (0.05)  &   6       \\
SN2013bz (SNhunt188)                              &   2MASX J13265081-1001263              &   $0.0192$ &   0.76 (0.02)        &  1.12 (0.05)   &  4,7     \\
SN2014dl (CSS140925:162946+083831)  &   UGC 10414                                         &   $0.0330$ &   1.05 (0.07)$^*$  &  1.22 (0.05)  &  3,4,8,9       \\
OGLE-2014-SN-107                                   &   APMUKS(BJ) B004021.02-650219.5  &   $0.0664$\tablenotemark{e} &   1.12 (0.05)         &  1.19 (0.05)  &   10      \\
ASASSN-14kd                                            &   2MASX J22532475+0447583             &   $0.0243$\tablenotemark{e} &   0.79 (0.01)         &  1.13 (0.05)  &  3,11       \\
OGLE-2014-SN-141                                   &   2MASX J05371898-7543157               &   $0.0625$\tablenotemark{e} &   0.62 (0.09)         &  1.24 (0.05)  &   12      \\
SN2014eg                                                   &   ESO 154-G 010                                  &   $0.0186$ &  0.92 (0.01)         &   1.17 (0.05)  &  3,4     \\
\enddata
\tablenotetext{a}{Heliocentric redshift are from the NASA/IPAC Extragalactic Database (NED) unless otherwise indicated.}
\tablenotetext{b}{\dm~decline rate in magnitudes \citep{phillips93} as measured with SNooPy. The 1$\sigma$ error is given in between parentheses.
The * symbol indicates values derived from template fits only.}
\tablenotetext{c}{\sbv~color stretch \citep{burns14} as measured with SNooPy templates. The 1$\sigma$ error is given in between parentheses.}
\tablenotetext{d}{References to spectroscopic observations.}
\tablenotetext{e}{Redshift measured by CSP-II.}
\tablerefs{
(1) \citet{scalzo14};
(2) \citet{tomasella13};
(3) This paper;
(4) PESSTO;
(5) \citet{Blagorodnova13};
(6) \citet{benitez-herrera13};
(7) \citet{ochner13};
(8) \citet{drake14};
(9) \citet{stahl20}
(10) \citet{takats14};
(11) \citet{firth14};
(12) \citet{dimitriadis14}
}
\end{deluxetable*}

\clearpage

\clearpage

\clearpage

\clearpage

\end{document}